\documentclass[reprint,twocolumn,aps,prl,
amsmath,amssymb,floatfix,groupedaddress,longbibliography]{revtex4-2}
\usepackage{float,array}
\usepackage{makecell, adjustbox,dsfont,tabularx}
\usepackage{graphicx}
\graphicspath{ {images/}, {figs/} }
\usepackage{tablefootnote}
\usepackage{todonotes}
\usepackage[normalem]{ulem}
\usepackage{booktabs}
\usepackage{dcolumn}
\usepackage{bm}

\usepackage[colorlinks=true,linktoc=all,urlcolor=cyan,citecolor=blue,linkcolor=blue]{hyperref}
\usepackage[capitalise]{cleveref}

\usepackage{physics,comment,mathtools}
\usepackage{etoolbox}

\usepackage[dvipsnames]{xcolor}
\usepackage{color}
\usepackage{colortbl}
\usepackage[caption=false]{subfig}
\makeatletter
\newcommand{\colorcaption}[2][]{%
  \begingroup%
  \renewcommand{\@caption@fignum@sep}{ (color online). }%
  \caption[#1]{#2}%
  \endgroup%
}
\newcommand{\tb}[1]{\textbf{#1}}

\usepackage{orcidlink}
\makeatother

\definecolor{profreplycolor}{RGB}{220,50,47} 

\begin{document}

\title{Interacting many-body non-Hermitian systems as Markov chains}
\author{Zichang Hao\,\orcidlink{0009-0009-3397-3443}}
\affiliation{Department of Physics, National University of Singapore, Singapore 117542}
\author{Wei Jie Chan\,\orcidlink{0000-0001-8310-0733
}}
\email{{wj\_chan@nus.edu.sg}}
\affiliation{Department of Physics, National University of Singapore, Singapore 117542}
\author{Ching Hua Lee\,\orcidlink{0000-0003-0690-3238}}
\email{{phylch@nus.edu.sg}}
\affiliation{Department of Physics, National University of Singapore, Singapore 117542}
\date{\today}

\begin{abstract}
Rich phenomenology emerges at the intersection of non-Hermiticity and many-body dynamics, yet physically realizable implementations remain challenging. 
In this work, we propose a general formalism that maps non-Hermitian many-body Hamiltonians to the Laplacians of Markov chains, such that wavefunction amplitudes are re-interpreted as stochastic many-body configuration probabilities. Despite explicitly preserving all state transition processes and inheriting analogous non-Hermitian localization and state-space fragmentation, our Markov chain processes exhibit distinct steady-state behavior independently of energetic considerations that govern quantum evolution. 
We demonstrate our framework with two contrasting representative scenarios, one involving asymmetric (biased) propagation with exclusion interactions, and the other involving flipping pairs of adjacent spins (agents). These results reveal robust and distinctive signatures of non-Hermitian phenomena in classical stochastic settings such as ecological and social networks, and provide a versatile framework for studying non-reciprocal many-body dynamics across and beyond physics.
\end{abstract}
\maketitle

\emph{Introduction}.---Non-Hermitian systems sparked great interest both experimentally \cite{zou2021observation,zhang2021observation,shen2025observation,koh2025interacting,zhao2025two,liu2018observation,shen2024enhanced,zou2024experimental,zhang2021observation,gu2022transient}  and theoretically \cite{ashida2020non, bergholtz2021exceptional,lee2016anomalous,gong2018topological,kawabata2019symmetry,yao2018edge,yokomizo2019non,yao2018non,okuma2020topological,lee2019anatomy,kunst2018biorthogonal,xiong2018does,song2019non,longhi2019probing,xiao2020non,helbig2020generalized,li2020critical,shen2022non,li2022non,shimomura2024general,qin2022non,jiang2023dimensional,yoshida2024non,arouca2020unconventional,lee2021many,wang2024non,bergholtz2021exceptional,lin2023topological,qin2024kinked,qin2023universal,liu2024non,yang2024non,gliozzi2024many,liu2024localization,Li2022c,Meng2025,Shen2024hyperbolic,lee2022exceptional,Li2025phase,Li2024exact,zhang2021tidal,fruchart2021non,hanai2019non,hanai2024nonreciprocal}.
While the single-particle regime is already largely well-understood, particularly for exceptional points \cite{jin2022exceptional,bergholtz2021exceptional,Li2023EPnanoscale,Zhao2024EPcircuits} and complex-deformed non-Hermitian skin effect (NHSE) bands \cite{yao2018edge,yokomizo2019non,lee2019anatomy,yao2018non,okuma2020topological,song2019non,li2020critical,yoshida2024non,Shen2024hyperbolic,lee2022exceptional}, 
many-body interacting non-Hermitian systems exhibit more exotic phenomena beyond these frameworks \cite{hamazaki2019non, nakagawa2018non, mu2020emergent, yoshida2024non, shimomura2024general, shen2022non,gliozzi2024many,lee2021many,qin2024occupation,Hamanaka2024,koh2025interacting,suthar2022non,Qin2025}, such as non-Hermitian skin clusters \cite{shen2025observation}, interacting cluster bursts \cite{koh2025interacting}, multifractality \cite{Hamanaka2024}, kondo effect \cite{nakagawa2018non} and occupation-dependent NHSE \cite{qin2024occupation}. 
Experimental realizations of such systems remain challenging, despite promising developments in ultracold atomic platforms and superconducting circuits \cite{zhang2021observation, liang2022dynmaics, wan2025non,kao2019introduction,zhao2025two,ren2022,Wang2024controllable,Ohnmacht2025,Chen2021quantumjumps}.

Interestingly, parallels of non-Hermitian many-body phenomena can be observed in various everyday-life settings, particularly in stochastic processes involving complex network dynamics \cite{merris1995survey, mirzaev2013laplacian, shirazi2009mapping,radicchi2018uncertainty}, population dynamics \cite{lotka1925elements,lotka1927fluctuations,volterra1927fluctuations,royama2012analytical,newman2014modelling,tuljapurkar2013population}, games \cite{goeree1999stochastic, traulsen2009stochastic, bertotti2004discrete}, and queues \cite{whitt2002stochastic, borovkov2012stochastic, kendall1953stochastic,nelson2013probability,kao2019introduction}. These systems involve asymmetric interactions that break detailed balance, leading to non-conservative dynamics analogous to those found in open quantum systems.
While some formalisms have captured intriguing new physics, these developments have, with few exceptions\cite{sawada2024role,nelson2024nonreciprocity,agudo2024topological}, been largely explored outside the traditional scope of condensed matter physics \cite{bak2013nature,frigg2003self,barabasi1999emergence,toner1998flocks,turcotte1999self,posfai2016network,boccaletti2006complex,castellano2009statistical,epstein2012generative,miller2009complex}.

Below, we present a general framework that maps quantum many-body Hamiltonians to Markov chain Laplacians that possess identical state transitions. It enables direct interpretations of many-body wavefunction amplitudes as configuration probabilities, connecting quantum dynamics to real-world processes \cite{merris1995survey, mirzaev2013laplacian, shirazi2009mapping,radicchi2018uncertainty,lotka1925elements,lotka1927fluctuations,volterra1927fluctuations,royama2012analytical,newman2014modelling,tuljapurkar2013population,goeree1999stochastic, traulsen2009stochastic, bertotti2004discrete,whitt2002stochastic, borovkov2012stochastic, kendall1953stochastic,nelson2013probability,kao2019introduction} with analogous non-Hermitian localization and state space fragmentation, but distinct steady-state behaviors. 

We illustrate our mapping with two representative setups. The first, an interacting Hatano-Nelson chain, maps onto a biased random walk with exclusion interactions that possess markedly different asymptotic dynamics. The second, which involves spin-flips among correlated adjacent spins, showcases how the particle number parity and ``Neel" order of the initial state can ultimately constraint the final steady-state, independently of energetic considerations from Hamiltonian dynamics.

\renewcommand{\arraystretch}{1.2}
\begin{table}
\resizebox{0.48\textwidth}{!}{     
\begin{tabular}{|c|c|c|}
\cline{1-3} 
\multicolumn{1}{|c|}{\textbf{}} & \textbf{Quantum systems}    & \textbf{Markov chains}  \\ 
\cline{1-3}
\textbf{Equation}&\begin{tabular}[c]{@{}c@{}}$\frac{d\Psi(t)}{dt}=-iH\Psi(t)$\\ $\sum_i |\psi_i|^2 =1$ if Hermitian\\ \end{tabular} & 
\begin{tabular}[c]{@{}c@{}}$\frac{d\Psi(t)}{dt}=-L\Psi(t)$\\ $\sum_i \psi_i =1$ always\\  \end{tabular}
  \\ 
 \cline{1-3}
\textbf{Probability}& $|\psi_i|^2$ with $ \psi_i \in \mathbb{C}$ &  $\psi_i$ with $\psi_i \in \mathbb{R}_{\geq 0}$
  \\ 
  \cline{1-3}
\textbf{\begin{tabular}[c]{@{}c@{}}Dynamics\end{tabular}}& $\Psi(t)=\exp\left(-iHt\right)\Psi(0)$ & \begin{tabular}[c]{@{}c@{}}$\Psi(t)= \exp\left(-L t\right) \Psi (0)$\end{tabular}  \\ 
\cline{1-3}
\textbf{Eigenvalues} & \begin{tabular}[c]{@{}l@{}}$\Re(E):$ oscillation\\ $\Im(E):$ decay\end{tabular} & \begin{tabular}[c]{@{}l@{}}$\Re(\omega):$ decay\\ $\Im(\omega):$ oscillation\end{tabular}  \\ 
\cline{1-3}
\end{tabular}}
\caption{\textbf{Distinction between quantum systems and their stochastic Markov chain analogs.} 
A quantum Hamiltonian $H$ and its analogous Markov chain Laplacian $L$ with similar state transitions exhibit markedly different physics due to different equations of motion and definitions of probability. In Markov chains, the state amplitude $\psi_i$ in $\Psi=\sum_i\psi_i|i\rangle$ are restricted to real and non-negative values that represent probabilities, and decay under real, not complex, eigenvalues. Unlike in quantum systems, total state occupancy probability is conserved by construction, even for non-Hermitian processes.   }
\label{tab:Tab1}
\end{table}

\begin{table*}[htbp]
\centering
\footnotesize
\renewcommand{\arraystretch}{1.2}
\setlength{\tabcolsep}{2pt}
\begin{tabular}{|c| c| c |c |c|}
\hline
& \textbf{Hamiltonian}   & \textbf{Laplacian} & \textbf{Particle hoppings/transitions}\\ 
\hline 
\makecell[c]{Interacting \\Hatano-Nelson\\ model} & $\begin{aligned}  
    H_{\lambda}=&\sum_{x}\sum_{\pm}\lambda_{\pm} \hat{b}_{x\pm1}^{\dagger}\hat{b}_{x} \\&\cdot(n_{max}-\hat{\rho}_{x\pm1}) 
\end{aligned}$                                   & 
$\begin{aligned} 
    L_{\lambda}=&\sum_{x}\sum_{\pm}\lambda_{\pm} (\hat{\rho}_{x}-\hat{b}_{x\pm1}^{\dagger}\hat{b}_{x})\\&\cdot(n_{max}-\hat{\rho}_{x\pm1})
\end{aligned}$                                   & 
\raisebox{-0.5\height}
{\includegraphics[width=0.25\textwidth]{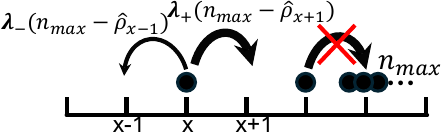}}                  \\
\hline 
\makecell[c]{Anti-correlated\\ spin-flip\\ model}                                 & $H_u=\sum_{x,\pm}u_{\pm}s_{x}^{\mp}s_{x+1}^{\pm}$     & 
$\begin{aligned}L_u= &\sum_{x,\eta=\{\pm\}}u_{\eta}\Big(\hat{n}_{x,\eta}\hat{n}_{x+1,-\eta}\\ &-\hat{b}_{x,-\eta}^{\dagger}\hat{b}_{x,\eta}\hat{b}_{x+1,\eta}^{\dagger}\hat{b}_{x+1,-\eta}\Big) \end{aligned}$          & 
\raisebox{-0.5\height}{\includegraphics[width=0.3\textwidth]{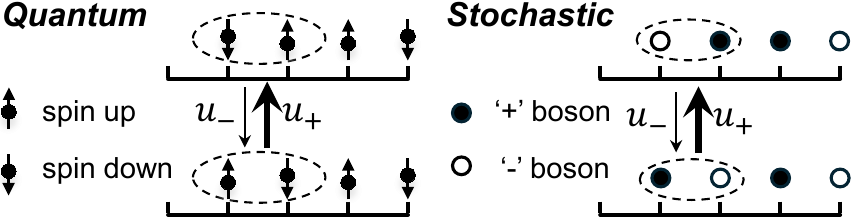}}                           \\
\hline 
\makecell[c]{Correlated \\ spin-flip   \\ model}                & $H_t=\sum_{x,\pm}t_{\pm}s_{x}^{\pm}s_{x+1}^{\pm}$      & $\begin{aligned}L_t= &\sum_{x,\eta=\{\pm\}}t_{\eta}\Big(\hat{n}_{x,-\eta}\hat{n}_{x+1,-\eta} \\&-\hat{b}_{x,\eta}^{\dagger}\hat{b}_{x,-\eta}\hat{b}_{x+1,\eta}^{\dagger}\hat{b}_{x+1,-\eta}\Big) \end{aligned}$                             & \raisebox{-0.5\height}{\includegraphics[width=0.3\textwidth]{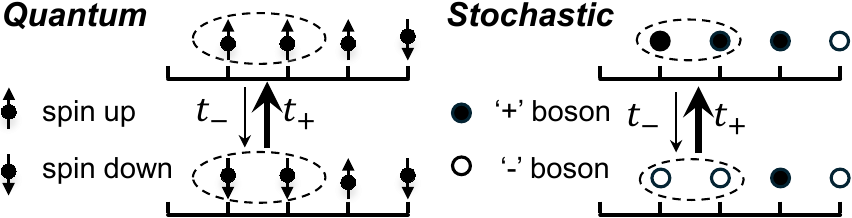}}\\
\hline
\end{tabular}
\caption{\textbf{Many-body Hamiltonians and their corresponding Markov chain Laplacians.}
(Top) The stochastic analog of the interacting Hatano-Nelson (HN) Hamiltonian $H_\lambda$ [\cref{eq:interact-HN-ham}] is the Markov chain Laplacian $L_\lambda$ [\cref{eq:interact-HN-lap}], both which possess unequal left/right hopping amplitudes $\lambda_\pm(n_{max}-\hat\rho_{x\pm 1})$ that linearly decreases with the occupancy of the destination site, which hence cannot exceed $n_{max}$.  
(Center) The model $H_u$, which flips adjacent unlike (anti-correlated) spins, maps onto a stochastic Laplacian $L_u$ [\cref{eq:spin-flip-lap}] that essentially reduces to the $L_\lambda$ above when $n_{max}=1$, since the spin flips can be interpreted as particle transfer.
(Bottom) However, the model $H_t$, which flips adjacent like (correlated) spins, maps onto a very different Laplacian $L_t$ [\cref{eq:pair-flip-lap}] whose steady state is dictated by the initial "Neel" order $m$ or particle number parity $\Pi$ [\cref{eq:even-odd-imbalance,eq:boson-number-parity}].
Here $\hat{\rho}_x = \hat{b}^{\dagger}_x\hat{b}_x$ is the particle number operator, and spin-up/down states are encoded using two bosonic species labeled by $\pm$, with $s_x^\pm = \hat{b}_{x,\pm}^\dagger \hat{b}_{x,\mp}$ flipping the spins. The on-site number operator $\hat{n}_{x,\pm}=\hat{b}_{x,\pm}^{\dagger}\hat{b}_{x,\pm}$ track the occupation of each species.
}
\label{tab:Tab2}
\end{table*}

\noindent{\emph{Markov chain formalism for many-body interactions}}.--- 
To interpret a quantum interaction as a stochastic Markov chain process, we first restrict the quantum many-body Hilbert space to the Markov chain's state space. 
For a Markov chain, a generic state $\Psi=\sum_{\vec n}\psi_{\vec n} |\vec n\rangle$ where $\sum_{\vec n}\psi_{\vec n}=1$, such that the coefficient $\psi_{\vec n}\geq 0$ represents the probability of being in the state $\vec n$; this is \emph{different} from  quantum mechanics where it is $|\psi_{\vec n}|^2$ that represents the probability [\cref{tab:Tab1}].  
To model many-body configurations, the basis is taken to be $|\vec n\rangle = |n_1,...,n_N\rangle$, which encodes having $n_x$ particles at site $x=1,2,...$, with 
\begin{equation}
\begin{aligned}
    \hat b_x^\dagger|...,n_x,...\rangle&= |...,n_x+1,...\rangle\\
    \hat b_x|...,n_x,...\rangle&= n_x|...,n_x-1,...\rangle.
\end{aligned}
\label{eq:crea-anni-operator}
\end{equation}
We have $|\vec n\rangle = (\hat b_1^\dagger)^{{n_1}} (\hat b_2^\dagger)^{{n_2}} \dots (\hat b_N^\dagger)^{{n_N}} |0\rangle$, since $\hat b_x|...,n_x,...\rangle=\hat b_x\hat b_x^\dagger|...n_x-1,...\rangle=(1+\hat{\rho}_x)|...n_x-1,...\rangle=n_x|...,n_x,...\rangle$, where the number operator is $\hat{\rho}_x = b^\dagger_x\hat b_x$. For ease of notation, we choose to normalize these basis states without the $\sqrt{\prod_x n_x!}$ factor, different from many texts.

We next relate any given quantum Hamiltonian $H$ with its Markov chain analog $L$. 
In a Markov process, the state evolves according to 
\begin{equation}
\Psi(t+\Delta t)= T_{\Delta t}\Psi(t)=\Psi(t)-L\Psi(t)\Delta t,
\label{eq:markov-state-evolve}
\end{equation}
where $T_{\Delta t}=\mathbb{I}-L\Delta t$ is the transition matrix, such that 
\begin{equation}
\frac{\mathrm{d}\Psi(t)}{\mathrm{d}t}=-L\Psi(t).
\label{eq:state-evolve}
\end{equation}
While this may superficially resemble the Schr\"odinger's equation with Wick-rotated time, we cannot simply replace $L$ by $iH$ due to the different probabilistic interpretation of the Markov chain state amplitude. Specifically, to implement the same hopping processes from a given $H$, we can define a corresponding Markov chain as
\begin{equation}
    \Psi(t+\Delta t) = \Psi(t) +(H-D)\Psi(t)\Delta t,
    \label{eq:state-evolve-H-to-L}
\end{equation}
where $D$ contains only on-site terms (is a diagonal matrix) introduced to ensure probability conservation \cite{merris1994laplacian,merris1995survey,mirzaev2013laplacian}. Comparing \cref{eq:state-evolve-H-to-L} and \cref{eq:state-evolve}, we identify $L=D-H$ as the Laplacian matrix~\footnote{This should not be confused with electrical circuit Laplacians \cite{imhof2018topolectrical,lee2018topolectrical,stegmaier2021topological,hofmann2019chiral,sahin2025topolectrical,stegmaier2024topological,helbig2020generalized} based differently on Kirchhoff's law, which can contain higher time derivatives from phase-shift elements such as capacitors and inductors.} of the state-space transition graph, since $D$ ensure that each column in $L=D-H$ sums to zero for probability conservation. Considerable freedom exists in the choice of $D$, since  all on-site terms conserve probability -- see \cref{tab:Tab2} for the illustrative models discussed later. 

\cref{tab:Tab1} summarizes various key differences between the quantum mechanical $H$ and its stochastic analog $L$, which hence necessarily exhibit distinct new physics despite inheriting the same state transitions (off-diagonal matrix elements).    
Notably, quantum interference, which relies on negative or complex state amplitudes, cannot occur in a Markov chain with $\psi_i\in\mathbb{R}_{\geq 0}$. 
Furthermore, due to the missing factor of $i$ in \cref{eq:markov-state-evolve}, the real and imaginary parts of the eigenvalues hold opposite roles in quantum vs. Markov chain dynamics.
Note that, due to probability conservation, a decaying Markov chain state amplitude (from real eigenvalues) must be accompanied by increases in the occupancy of other states, even if the state transitions are non-Hermitian.

\noindent{\emph{HN model with exclusion interactions}}.--- 
As the first example, we generalize the Hatano-Nelson (HN) model \cite{hatano1996localization} to have occupancy-dependent asymmetric hopping interactions, and show how it maps to a stochastic biased random walk model with exclusion~[\cref{tab:Tab2}]. 
The Hamiltonian is
\begin{equation}
H_{\lambda}=\sum_{x}\sum_{\pm}\lambda_{\pm} \hat{b}_{x\pm1}^{\dagger}\hat{b}_{x}(n_{max}-\hat{\rho}_{x\pm1}) ,
 \label{eq:interact-HN-ham}
\end{equation}
where $\lambda_{\pm}$ represents the rightward/leftward hopping amplitude.
The system exhibits non-reciprocal hopping dynamics whenever $\lambda_+\neq\lambda_-$. 
The density-density and density-hopping 4-operator  $\hat{b}_{x\pm1}^{\dagger} \hat{b}_{x} (n_{max} - \hat{\rho}_{x\pm1})$ introduces an exclusion interaction that enforces a maximum occupancy constraint of $n_{max}$ particles per site. 
Keeping the same hoppings as in \cref{eq:interact-HN-ham}, we write down a Markov chain process on a non-reciprocal $1$D spinless lattice with a monoatomic unit cell and nearest neighbor jumps:
\begin{equation}
    \begin{aligned}
    \Psi_{\vec n}(t+\Delta t) &= \Psi_{\vec n}(t)+ \sum_{x,\pm} \left(\hat b^\dagger_{x\pm 1}\hat b_x -\hat \rho_x\right)\\
    &\cdot(n_{max}-\hat{\rho}_{x\pm1})\Psi_{\vec n}(t)\Delta t.
    \end{aligned}
    \label{eq:interact-HN-evo}
\end{equation}
The overall occupation probability is conserved across the many-body basis states, since $\left(\hat b^\dagger_{x+1} \hat b_x - \hat \rho_x\right)|n_x, n_{x+1}\rangle=\left(\hat b^\dagger_{x+1}- \hat b^\dagger_{x}\right) \hat b_x | n_x, n_{x+1}\rangle
=|n_x-1, n_{x+1}+1\rangle-| n_x, n_{x+1}\rangle$, such that any increase in the new configuration is compensated by an equal decrease in the old one. 
Comparing \cref{eq:interact-HN-evo} with \cref{eq:markov-state-evolve}, the Laplacian governing the stochastic analog of $H_\lambda$ [\cref{eq:interact-HN-ham}] is
\begin{equation}
L_{\lambda}=\sum_{x,\pm} \lambda_\pm\left(\hat \rho_x-\hat b^\dagger_{x\pm 1}\hat b_x \right)(n_{max}- \hat \rho_{x\pm 1}),
\label{eq:interact-HN-lap}
\end{equation}
where $(n_{max} - \hat \rho_{x\pm 1})$ imposes a soft constraint at $n_{max}$, disallowing any particle into site $x\pm1$ when it has already been occupied by $n_{max}\in \mathbb{Z}$ particles. \cref{eq:interact-HN-lap} is mathematically equivalent to the $K$-exclusion process—a generalization of the asymmetric simple exclusion process (ASEP) \cite{schutz1994non, cocozza1985processus, evans2014condensation, ayyer2024exactly}, and in general describes constrained directional agent propagation from the interplay between the NHSE (biased directed accumulation) and inter-agent repulsion, encountered in ion transport through narrow biological channels or traffic flow in bottle-necked lanes \cite{Schreckenberg1995discrete,kaufman2015coulomb} [see S7 in \cite{mysupp} for more examples].

\begin{figure}
    \centering
    \includegraphics[width=\linewidth]{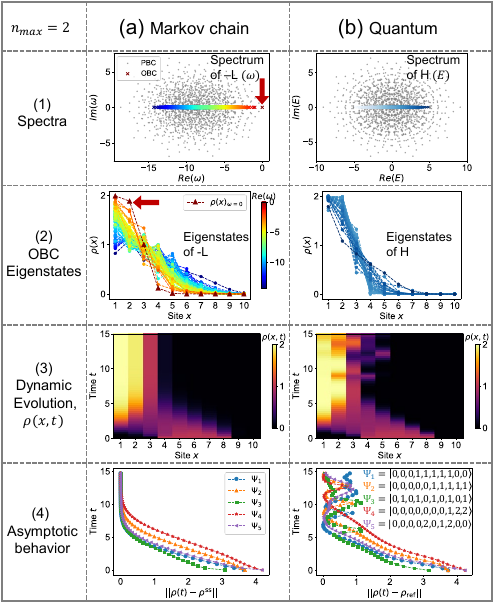}
    \caption{\tb{Comparison between stochastic (Markov chain) and quantum (Hamiltonian) formulations of the Hatano-Nelson model with exclusion interactions. } 
    We consider a half-filled system with asymmetric hoppings $\lambda_+ = 0.1$, $\lambda_- = 1$ among $N = 10$ sites, with 
    occupation limited to $n_{max} = 2$ particles per site. 
    \tb{(1)} Laplacian ($-L_\lambda$ [\cref{eq:interact-HN-lap}]) and Hamiltonian ($H_\lambda$ [\cref{eq:interact-HN-ham}]) eigenspectra, both of which are real under OBCs, but not PBCs. A zero mode (red arrow) always exists as the Markov chain steady-state.
    \tb{(2)} Laplacian ($-L_\lambda$) and Hamiltonian ($H_\lambda$) eigenstate densities $\rho(x)$ under OBCs. The $\lambda=0$ Markov chain steady state density profile $\rho^{ss}=\rho(x)_{\omega=0}$ (red arrow) is Fermi-Dirac-like,  as are most Hamiltonian eigenstates.
    \tb{(3)} Dynamical state evolution from a given initial state $\Psi_1 = \ket{0,0,0,1,1,1,1,1,0,0}$ under OBCs. The system relaxes to the steady state profile $\rho^{ss}$ under Markov chain evolution (Left), but exhibits persistent oscillations under quantum evolution (Right). 
    \tb{(4)} Asymptotic dynamical behavior: For the Markov chain (Left), $\rho(x)$ universally converges to the unique OBC steady state profile $\rho^{ss}$ irrespective of the initial state, as indicated. In contrast, under quantum evolution, $\rho(x)$ oscillates indefinitely around a reference state density $\rho_{\text{ref}} = [2, 2, 1, \dots, 0, 0]$ which represents a maximally left-localized configuration consistent with $n_{max} = 2$. 
    }
    \label{fig:Fig1}
\end{figure}

One can show [see S1 in \cite{mysupp}] that the expected steady-state particle density $\rho^{ss}(x) = \langle \hat{\rho}_x \rangle$ under open boundary conditions (OBCs) closely resembles a rescaled Fermi-Dirac distribution
\begin{equation}
    \rho^{ss}(x)=\frac{n_{max}}{1+e^{ \frac{x-n/n_{max}}{k_B T_{\text{eff}}} } }, \quad k_{B} T_{\text{eff}} = \frac{1}{\ln(\lambda_-/\lambda_+)},
    \label{eq:fermi-like-distribution}
\end{equation}
where $n$ denotes the total particle number. This behavior is illustrated in \cref{fig:Fig1}(a1–a2), as indicated by the red arrow.
Other eigenstates of $L$ do not exhibit a similar profile, but they are irrelevant to the asymptotic time dynamics, since $\Psi(t)=e^{-Lt}\Psi(0)$ [see \cref{tab:Tab1}] for Markov chains, such that all initial states converge towards the unique zero mode ($\omega=0$) steady state density distribution, i.e., $\rho^{ss}=\rho(x)_{\omega=0}$[\cref{fig:Fig1}(a3-4)]. 
It is interesting that it assumes the same form as the Fermi-Dirac distribution, which arises as a variational solution in statistical mechanical ensembles (albeit in real space, not energy space).

In contrast, \emph{most} OBC eigenstates in the quantum interacting HN system exhibit such Fermi-Dirac-like profiles \cite{mu2020emergent,shen2025observation,shimomura2024general,shen2025observation,Cao2023Many} [\cref{fig:Fig1}(b1-2)], as can be deduced from many-body point-gap topology \cite{ kawabata2022many}, 
or directly from the Slater determinant in the case of fermions \cite{shen2025observation}. 
However, since $\Psi(t)=e^{-i Ht}\Psi(0)$ and the eigenenergies are real, no one eigenstate dominates the time evolution, and the dynamical states oscillate perpetually without settling down into any particular "Fermi skin" profile [\cref{fig:Fig1} (b3-4)].

\begin{figure*}
    \centering
    \includegraphics[width=\linewidth]{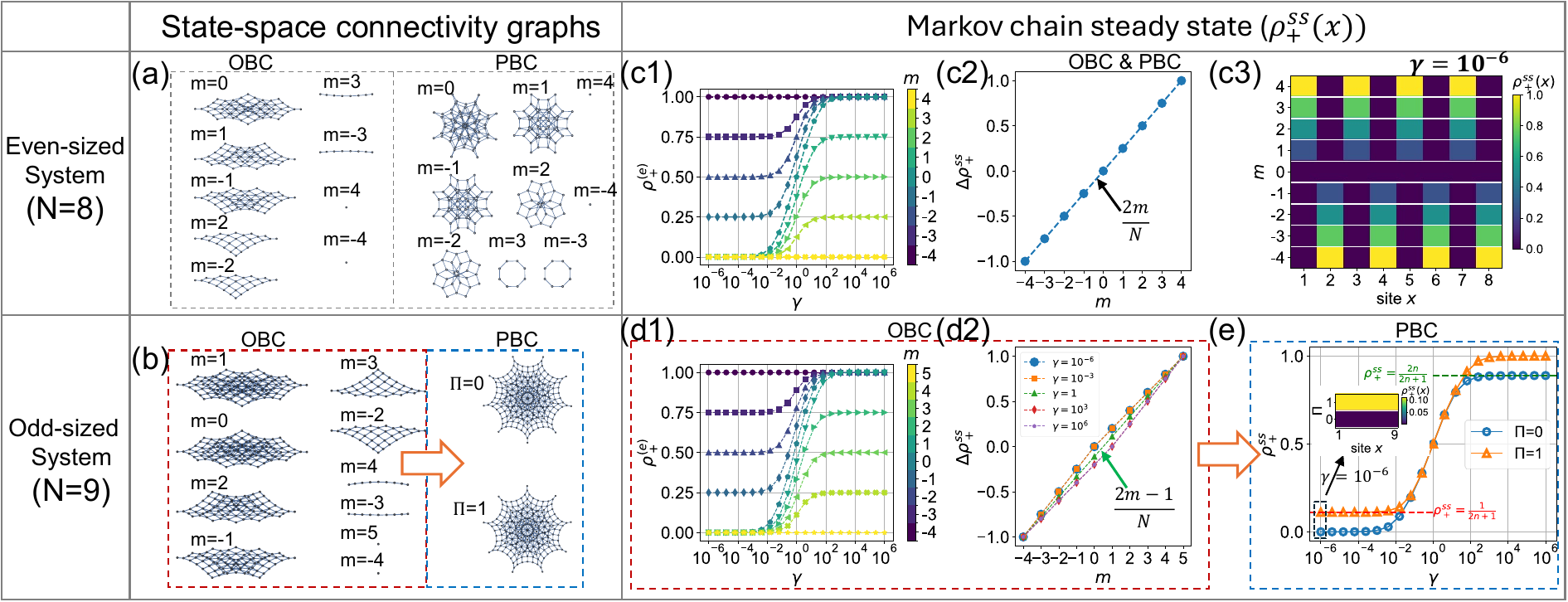}
    \caption{\tb{State space fragmentation and steady‐state profiles of the correlated spin-flip Laplacian [\cref{eq:pair-flip-lap}].}
    \tb{(a,b)} State space fragmentation structure: For even system sizes $N$, there are always $N+1$ fragments indexed by the odd-even site imbalance $m$ [\cref{eq:even-odd-imbalance}], and boundary conditions (OBC/PBC) only affect the internal structure of each subspace; for odd $N$, these $N+1$ subspace fragments remain under OBCs but not PBCs, where they are instead indexed by $\Pi=0,1$ [\cref{eq:boson-number-parity}].
    \tb{(c1-c3)} For even-sized systems, steady states universally exhibit staggered spatial profiles: $\rho_+^{(e)}$ at even sites and $\rho_+^{(o)}=\rho_+^{(e)}+\Delta\rho_+^{ss}$ at odd sites, as illustrated in (c3) with $\gamma=10^{-6}$. (c1) $\rho_+^{(e)}$ depends sharply on $\gamma = t_+/t_-$ and is almost quantized by $m$ at very small or large $\gamma$; (c2) The ``Neel" amplitude $\Delta\rho_+^{ss}$ increases linearly with $m$ (c2).  
    \tb{(d1,d2)} Odd-sized OBC systems exhibit qualitatively similar staggered steady-states as in (c), except that  $\Delta\rho_+^{ss}$ becomes non-linear in $m$ unless in the Hermitian $\gamma=1$ limit. 
    \tb{(e)} Even-$N$ PBC steady-states are spatially uniform (inset), with density $\rho_+^{ss}$ determined by the particle number parity $\Pi =0,1$ [\cref{eq:boson-number-parity}]; for even but not odd $\Pi$, $\rho_+^{ss}$ never approaches unity no matter how strong $\gamma$ is. 
    } 
    \label{fig:Fig2}
\end{figure*}

\noindent{\emph{Stochastic analogs of spin-flip models}}.--- 
We next examine spins systems with spin-flipping terms between adjacent spins, which occur naturally in various quantum materials \cite{Gulacsi1993,Jayaprakash1989} and interacting qubit chains \cite{Senko2015,Zhang2007}. There are two distinct possibilities $H_u$ and $H_t$ [see \cref{tab:Tab2}]: they act pairwise on two adjacent spins that are initially oriented oppositely/parallel to each other. 

The first possibility $H_u=\sum_{x,\pm} u_{\pm}s_x^{\mp}s_{x+1}^{\pm}$, with $s_x^\pm=\hat b^\dagger_{x,\pm}\hat b_{x,\mp}$, turns out to trivially reduce to the above ASEP stochastic analog $L_\lambda$ with $n_{max}=1$. From
\begin{equation}
     L_u =\sum_{x,\eta}u_{\eta}(\hat{n}_{x,\eta}\hat{n}_{x+1,-\eta}-\hat{b}_{x,-\eta}^{\dagger}\hat{b}_{x,\eta}\hat{b}_{x+1,\eta}^{\dagger}\hat{b}_{x+1,-\eta}),
     \label{eq:spin-flip-lap}
\end{equation}
imposing a maximal site occupancy of $n_{x,+}+n_{x,-} = 1$, gives $L_\lambda$ [\cref{eq:interact-HN-lap}]~\cite{mysupp}. 
Specifically, $n_{x,+}$, the + particle occupancy in its steady state, follows \cref{eq:fermi-like-distribution} with $n_{max}=1$, $\hat \rho_x \rightarrow \hat n_{x,+}$, $\lambda_\pm \rightarrow u_{\pm}$ and $\hat b_x \rightarrow \hat b_{x,+}$. Intuitively, that is because a single spin flip corresponds to toggling between the presence and absence of a particle, such that two adjacent flips between unlike spin states corresponds to the adjacent hopping of that particle.

However, the other possibility $H_t=\sum_{x,\pm} t_{\pm}s_x^{\pm}s_{x+1}^{\pm}=\sum_{x,\eta=\{\pm\}}t_{\eta}\hat{b}_{x,\eta}^{\dagger}\hat{b}_{x,-\eta}\hat{b}_{x+1,\eta}^{\dagger}\hat{b}_{x+1,-\eta}$, which flips correlated (aligned) adjacent spins together, behaves non-trivially in its stochastic interpretation, governed by the Markov chain Laplacian $L_t =D-H_t$ which takes the form 
\begin{equation}
    L_t =\sum_{x,\eta=\{\pm\}}t_{\eta}(\hat{n}_{x,-\eta}\hat{n}_{x+1,-\eta}-\hat{b}_{x,\eta}^{\dagger}\hat{b}_{x,-\eta}\hat{b}_{x+1,\eta}^{\dagger}\hat{b}_{x+1,-\eta}),
    \label{eq:pair-flip-lap}
\end{equation}
where $D=\sum_{x,\eta=\{\pm\}}t_{\eta}\left(\hat{n}_{x,-\eta}\hat{n}_{x+1,-\eta}\right)$ enforces overall probability conservation. 
With $n_{max}=1$, its basis states resemble fermionic Fock states e.g. $\ket{0,1,1,0,\dots}$, but are symmetric under exchange.

The correlated spin-flip term $\hat{b}_{x,\eta}^\dagger \hat{b}_{x,-\eta} \hat{b}_{x+1,\eta}^\dagger \hat{b}_{x+1,-\eta}$ enforces a local kinetic constraint, permitting only spin flips in adjacent pairs [\cref{tab:Tab2}]. While total spin (or particle number) is not conserved, what is conserved is the \emph{odd-even imbalance} i.e. ``Neel order parameter"~\cite{michael2015observation} 
\begin{equation}
    m =\sum_{x=1}^{N} (-1)^{x+1} n_{x,+} =\sum_{x=odd}n_{x,+}-\sum_{x=even}n_{x,+},
\label{eq:even-odd-imbalance}
\end{equation} 
which quantifies the imbalance of $+$ particles between the odd and even lattice sites, except when the system has odd length $N$ under periodic boundary conditions (PBCs), since an odd-length ring is not bipartite. As each $n_{x,+}$ contributes $\pm 1$ to $m$, the invariant $m$ can take $N+1$ different integer values. 
Hence, the many-body state space fragmentates into $N+1$ dynamically disconnected sectors labeled by $m$~\cite{moudgalya2022quantum, sala2020ergodicity}, as explicitly illustrated in \cref{fig:Fig2}(a,b). 

While such state space fragmentation occurs in both quantum and stochastic settings, only in the latter do there exist steady-state profiles that depend prominently on $m$ [\cref{fig:Fig2}(c)]. Interestingly, the steady-state occupation profile of the $+$ particles, defined as $\rho_+^{ss}(x) = \langle n_{x,+} \rangle$, exhibits an emergent staggered profile characterized by different (constant) values at odd and even sites, as illustrated in \cref{fig:Fig2}(c3): 
\begin{equation}
\rho_+^{ss}(x)=
\begin{cases}
   \rho_+^{(e)} & \text{if }x \text{ is even,} \\
   \rho_+^{(e)}+\Delta\rho_+^{ss} & \text{if }x \text{ is odd,} 
\end{cases}
\end{equation}
where
\begin{equation}
    \rho_+^{(e)}=\frac{\sum_{i=0}^{l-m}\binom{l}{i+m}\binom{l}{i} \gamma^{i} \cdot \frac{i}{l}  }{\sum_{i=0}^{l-m}\binom{l}{i+m}\binom{l}{i} \gamma^{i}}
\label{eq:pair-flip-model-ss} 
\end{equation}
for even-sized systems, $\gamma = \frac{t_+}{t_-}$ is the pair-flipping asymmetry, and $l=\lfloor N/2\rfloor$ [see S5 in \cite{mysupp} for analytical derivations, and S6 for $\rho_+^{(e)}$ for odd system sizes].
While it is intuitively expected that $\rho_+^{ss}(x)$ increases with $\gamma = \frac{t_+}{t_-}$, since the $+$ particles survive longer when $t_+>t_-$, what is intriguing is its staggered spatial profile. At large $|m|$, the propensity for finding a $+$ particle depends strongly on the odd/evenness of the site position, even though the Markov chain is manifestly translation invariant.

That such spatial ``Neel" inhomogeneity $\Delta \rho_+^{ss}$ exists can be traced to the dynamical invariance of $m$ which, once imprinted in the initial state, continues to dictate its eventual profile after reaching steady-state, spontaneously breaking its even-odd site symmetry. These even-odd amplitude fluctuations increase linearly with $m$:
\begin{equation}
    \Delta \rho_+^{ss} = \frac{2m}{N}
\end{equation}
for even system sizes $N = 2l$ [\cref{fig:Fig2}(c2)]. For odd $N$, $ \Delta \rho_+^{ss}$ exhibits more complicated dependence on $m$ [\cref{fig:Fig2}(d1-d2)] due to the different number of odd and even sites, as derived analytically in S5 of \cite{mysupp}. 

Notably, the boundary conditions only affect the fragmentation structure significantly when the system size $N$ is odd -- when switching from OBCs to PBCs, the number of state space fragments drops sharply, from $N+1$ to two [\cref{fig:Fig2}(b)].  
Under PBCs, the even-odd distinction between the sites and hence $m$ [\cref{eq:even-odd-imbalance}] becomes ill-defined, and the dynamics preserve only a $\mathbb{Z}_2$ index [\cref{fig:Fig2}(b)], characterized by the \emph{particle-number parity} of the $+$ species:
\begin{equation}
\Pi = \bigg(\sum_x n_{x,+}\bigg) \bmod 2 \in \{0,1\}. 
\label{eq:boson-number-parity}
\end{equation}
Accordingly, the steady-state PBC and OBC profiles are very different, despite the \emph{absence} of spatial non-Hermitian skin accumulation. Under PBCs, the two values of $\Pi \in \{0,1\}$  leads to two possible uniform steady-state densities [\cref{fig:Fig2}(e)] -- interestingly, for even but not odd $\Pi$, the density cannot approach $1$ no matter how small or large the asymmetry $\gamma$ is [see S6 in \cite{mysupp}]. 

Such steady-state behavior can manifest in real-world opinion dynamics, where agents each hold a binary opinion $n_{i,+}=\pm 1$. When two adjacent agents share the same opinion ($ n_{i,\pm} = n_{i+1,\pm} $), they either reinforce it with probability $ t_{\pm} $, or both switch with probability $ 1 - t_{\pm} $, as 
prescribed by $ \hat{b}_{x,\eta}^{\dagger} \hat{b}_{x,-\eta} \hat{b}_{x+1,\eta}^{\dagger} \hat{b}_{x+1,-\eta}$. The invariant $m$ represents fundamentally irremovable aspects of the collective opinion that manifests as inevitable biases based on the agents' position, despite all being like-minded with similar persuasive power.

\noindent\emph{Discussion.---} We present a Markov chain framework that gives alternative interpretations of non-Hermitian many-body Hamiltonians as memoryless stochastic processes involving multiple agents. Even though both interpretations stipulate identical state transitions, they differ greatly in terms of phenomenology, with different propensities for interference and entanglement, as well as dynamical evolution. 
Within this formalism, we explore two representative models, one originally harboring the NHSE with exclusion interactions, and the other with spin-flipping asymmetry. The resulting Markov chain dynamics exhibit steady states not reached in their quantum analogs, with emergent Fermi-Dirac-like and even-odd staggering profiles that lend insights to real-world processes such as opinion dynamics \cite{french1956formal,degroot1974reaching,sznajd2000opinion,castellano2009statistical}.

\begin{acknowledgments}
\noindent{\emph{Acknowledgements.---}} ZH would like to thank Ruizhe Shen for helpful discussions.  CHL acknowledges support from the Ministry of Education, Singapore (MOE Tier-II award numbers: MOE-T2EP50222-0003 and MOE-T2EP50224-0021) and (Tier-I WBS number: A-8002656-00-00). 
\end{acknowledgments}

\bibliography{references}
\clearpage
\setcounter{equation}{0}
\setcounter{figure}{0}
\setcounter{table}{0}
\setcounter{section}{0}
\setcounter{subsection}{0}
\setcounter{secnumdepth}{3}

\renewcommand{\thesection}{S\arabic{section}}
\renewcommand{\thesubsection}{S\arabic{section}.\arabic{subsection}}
\renewcommand{\thesubsubsection}{S\arabic{section}.\arabic{subsection}.\arabic{subsubsection}}
\makeatletter
\renewcommand{\p@section}{}          
\renewcommand{\p@subsection}{}       
\renewcommand{\p@subsubsection}{}    
\makeatother
\renewcommand{\theequation}{S\arabic{equation}}
\renewcommand{\thefigure}{S\arabic{figure}}
\renewcommand{\thetable}{S\arabic{table}}
\renewcommand{\theHtable}{S\arabic{table}} 

\crefname{section}{Sec.}{Secs.}
\crefname{subsection}{Sec.}{Secs.}
\onecolumngrid
\flushbottom
\setcounter{page}{1}
\begin{center}
\textbf{\large Supplemental Online Material for “Interacting many-body non-Hermitian processes as Markov chains” }
\end{center}
\date{\today}

\begin{center}
 {\small Zichang Hao, Wei Jie Chan, and Ching Hua Lee}  
\end{center}
\begin{center}
{\sl \footnotesize
Department of Physics, National University of Singapore, Singapore 117542
}
\end{center}
\begin{quote}
	{\small
		This supplementary material contains:\\
        (\ref{sec:s1}) derivations of the Fermi-Dirac-like steady-state distribution in the interacting HN model; \\
        (\ref{sec:s2}) demonstration of a formal equivalence between the interacting HN model and the K-exclusion process (K-ASEP); \\
        (\ref{sec:s3}) details about the construction of the Laplacian for spin models; \\
        (\ref{sec:s4}) the correspondence between the anti-correlated spin-flip model and the interacting HN model;  \\
        (\ref{sec:s5}) derivations of steady states for the correlated spin-flip    model with an even number of sites;\\
        (\ref{sec:s6}) derivations of steady states for the correlated spin-flip    model with an odd number of sites;\\
	(\ref{sec:s7}) physical interpretation of Laplacians.
	}   
\end{quote}

\section{Fermi-Like Steady State profile in the interacting HN model}\label{sec:s1}

In this section, we present an analytical derivation that shows how the steady state typically assumes a Fermi-like steady state profile in the stochastic Hatano-Nelson (HN) model with repulsive interactions that prevent multiple occupancy, under certain approximations. The derivation can be approached from two perspectives in \cref{sec:s1.1,sec:s1.3}, with the latter [\cref{sec:s1.3}] offering a practical advantage by avoiding truncation errors from Taylor expansion and thus often being the preferred approach. We also discuss the validity regime of these approximations.
\subsection{Fermi-like Steady State}\label{sec:s1.1}
We consider a Markov chain defined on a non-reciprocal one-dimensional lattice of size $N$, with a monoatomic unit cell, nearest-neighbor interactions, and $n$ particles. The evolution of the Markov state $\Psi(t)$, which represents the particle probability distribution, is governed by
\begin{gather}
    \Psi(t + \Delta t) = \left[\mathbb{I} - L_{\lambda} \Delta t\right] \Psi(t), \quad 
    \frac{\mathrm{d}\Psi(t)}{\mathrm{d}t} = -L_{\lambda} \Psi(t), 
    \label{eq:supp-markov-evolve} \\
    L_{\lambda} = \sum_{x} \sum_{\pm} \lambda_{\pm} \left( \hat{\rho}_{x} - \hat{b}_{x \pm 1}^{\dagger} \hat{b}_{x} \right) (n_{\mathrm{max}} - \hat{\rho}_{x \pm 1}),
    \label{eq:supp-interact-HN-Lap}
\end{gather}
where $L_{\lambda}$ is the Laplacian operator encoding the transition rates between state configurations. The operators $\hat{b}_{x}^{\dagger}$ and $\hat{b}_{x}$ denote bosonic creation and annihilation operators at site $x$, and $\hat{\rho}_{x} = \hat{b}_{x}^{\dagger} \hat{b}_{x}$ is the local occupation number operator. The factors $(n_{max} - \hat{\rho}_{x \pm 1})$ impose upper bounds on local occupation, prohibiting any further hopping of particles to sites $x\pm 1$ if there are already $n_{max}$ particles there. 
Hence the state space is spanned by Fock states $\ket{\vec{n}} = \ket{n_1, \dots, n_N}$, where $n_x \in \{0, \dots, n_{max}\}$ denotes the number of particles at site $x$ and $n_{max}$ is the maximum occupancy limit per site. We denote the state of such a processes at time $t$ as $\Psi(t) = \sum_{\vec{n}} \psi_{\vec{n}}(t) \ket{\vec{n}},$ with $\psi_{\vec{n}}(t)$ representing the probability associated with the configuration $\ket{\vec{n}}$ at time $t$.  

The Fock basis states $\ket{\vec{n}_i}$ are labeled by index $i = 1, \dots, \mathcal{D}$, with $\mathcal{D}$ denoting the dimension of the whole state space.
Each basis state is specified as $\ket{\vec{n}_i} = \ket{n_i(1), \dots, n_i(x), n_i(x+1), \dots, n_i(N)}$, where $n_i(x)$ denotes the occupation number at site $x$ in the $i$-th Fock state.
According to the dynamical equation [\cref{eq:supp-interact-HN-Lap}], 
the term $\lambda_- \hat{b}^{\dagger}_{x-1}\hat{b}_x (n_{max}-\hat\rho_{x-1})$ represents the probability of a transition from the $i$-th basis state $|\overrightarrow{n_{i}}\rangle=|n_{i}(1),\cdots,n_{i}(x),n_{i}(x+1),\cdots,n_{i}(N)\rangle$
to the state $
|n_{i}(1),\cdots,n_{i}(x)+1,n_{i}(x+1)-1,\cdots,n_{i}(N)\rangle$, where a particle moves from site $x+1$ to site $x$.
After a long time, the system reaches a \emph{steady state} $\Psi(t)=\sum_{i}\psi_{i}(t)|\overrightarrow{n_{i}}\rangle$ in which the probabilities $\psi_{i}(t)$ no longer change.
We define the average particle density at site $x$ as 
\begin{equation}
\rho(x)=\ev{\hat\rho_x}=\sum_{i=1}^{\mathcal{D}}\psi_{i}n_{i}(x),
\end{equation} where the average is taken over all possible occupancy configurations $i$, and $\mathcal{D}$ denotes the dimension of the state space.

\noindent\paragraph{Steady State Condition.} In the steady state, the net change to $\rho(x)$, i.e., $\Delta \rho(x)$, must be zero when considering all possible transitions between different occupancy configurations. This condition ensures that the total probability of a boson jumping into site $x$ equals the total probability of a boson jumping out of site $x$ to other sites. Consequently, $L_\lambda\Psi(t)=0$ (see \cref{eq:supp-markov-evolve}) gives the following discrete difference equation:
\begin{equation}
\begin{aligned}
&\lambda_{-}\sum_{i}\psi_{i}n_{i}(x+1)(n_{max}-n_{i}(x))+\lambda_{+}\sum_{i}\psi_{i}n_{i}(x-1)(n_{max}-n_{i}(x))=\\
&\lambda_{-}\sum_{i}\psi_{i}n_{i}(x)(n_{max}-n_{i}(x-1))+\lambda_{+}\sum_{i}\psi_{i}n_{i}(x)(n_{max}-n_{i}(x+1)).
\label{eq:supp-interacting-HN-ss-condition}
\end{aligned}
\end{equation}
By applying the mean-field approximation (see \cref{sec:s1.2}):
\begin{equation}
    \lambda_{\pm}\sum_{i}\psi_{i}n_{i}(x)(n_{max}-n_{i}(x\pm1))\approx \lambda_{\pm}\rho(x)(n_{max}-\rho(x\pm1)),
    \label{eq:supp-FD-approx}
\end{equation}
the steady state condition \cref{eq:supp-interacting-HN-ss-condition} is reformulated into 
\begin{equation}
    \Delta \rho(x)\approx
    \lambda_{-}\rho(x+1)(n_{max}-\rho(x))+\lambda_{+}\rho(x-1)(n_{max}-\rho(x))-\Big(\lambda_{-}\rho(x)(n_{max}-\rho(x-1))+\lambda_{+}\rho(x)(n_{max}-\rho(x+1))\Big)=0.
    \label{eq:supp-approx-diff-eq}
\end{equation}
Simplifying \cref{eq:supp-approx-diff-eq} by Taylor expanding $\rho(x\pm1) = \rho(x) \pm \rho'(x) + \rho''(x)/2$, we have
\begin{equation}
    \rho''(x) - \frac{4(\lambda_{-}-\lambda_{+})}{(\lambda_{-}+\lambda_{+})n_{max}} \rho'(x)\rho(x) + \frac{2(\lambda_{-}-\lambda_{+})}{\lambda_{-}+\lambda_{+}} \rho'(x) =0.
\end{equation} 
Noting that $\rho''(x)=\frac{d\rho'(x)}{dx} = \frac{d\rho'(x)}{d\rho}\,\rho'(x),$
we substitute this into the equation and factor out $\rho'(x)$ (assuming $\rho'(x)\neq 0$) to obtain
\begin{equation}
    \frac{d\rho'}{d\rho} - \frac{4(\lambda_{-}-\lambda_{+})}{(\lambda_{-}+\lambda_{+})\,n_{max}}\,\rho + \frac{2(\lambda_{-}-\lambda_{+})}{\lambda_{-}+\lambda_{+}} = 0.
\end{equation}
Integrating both sides with respect to $\rho$, we have
\begin{equation}
    \rho'(x) = \frac{2(\lambda_{-}-\lambda_{+})}{(\lambda_{-}+\lambda_{+})\,n_{max}}\,\rho(x)^2 - \frac{2(\lambda_{-}-\lambda_{+})}{\lambda_{-}+\lambda_{+}}\,\rho(x) + C_1,
    \label{eq:supp-one-order-diff}
\end{equation}
where $C_1$ is an integration constant.
This can be put into separable form
\begin{equation}
    \frac{d\rho}{\displaystyle \frac{2(\lambda_{-}-\lambda_{+})}{(\lambda_{-}+\lambda_{+})\,n_{max}}\,\rho^2 - \frac{2(\lambda_{-}-\lambda_{+})}{\lambda_{-}+\lambda_{+}}\,\rho + C_1} = dx.
    \label{eq:supp-seperable-form}
\end{equation}
Integrating both sides of \cref{eq:supp-seperable-form}, we obtain
\begin{equation}
    \frac{1}{\mathcal{M}}\int \frac{d\rho}{\left(\rho-\frac{n_{max}}{2}\right)^2+\alpha^2} = x + C_2,
\end{equation}
where we have defined
\begin{equation}
\mathcal{M}=\frac{2(\lambda_{-}-\lambda_{+})}{(\lambda_{-}+\lambda_{+})\,n_{max}}, \qquad  \alpha^2=\frac{4C_1-An^2_{max}}{4A^2},
\label{eq:supp-derivation-const}
\end{equation} and $C_2$ is an integration constant.
Solving the integral, we obtain the general solution:
\begin{equation}
    \rho(x)=\frac{n_{max}}{2} + \alpha\,\tan\Bigl[\mathcal{M}\alpha\,(x+C_2)\Bigr].
    \label{eq:supp-HN-rhox}
\end{equation}

\noindent\paragraph{Fermi-Like Distribution.} 
We demonstrate that under open boundary conditions (OBCs) and in the thermodynamic limit, e.g., $N\to \infty$ and $n/N=const$, $\rho(x)$ obeys the Fermi-Dirac (FD) distribution. 
Assuming $\lambda_- >\lambda_+$ without loss of generality, particle accumulation at the left boundary i.e., $\rho(0) = n_{max}$ with $\rho'(0)=0$ is expected due to the presence of non-reciprocal hopping and a finite occupation limit $n_{max}$. Substituting this condition 
into \cref{eq:supp-one-order-diff}, we obtain $C_1=0$, leading to $\alpha=i n_{max}/2$ [\cref{eq:supp-derivation-const}]. With $\alpha=i n_{max}/2$,  we further simplify \cref{eq:supp-HN-rhox}:
\begin{align}
\rho(x)
    &=\frac{n_{max}}{2} - \frac{n_{max}}{2} \tanh\Bigl[\mathcal{M}\frac{n_{max}}{2}\,(x+C_2)\Bigr]\\
    &=\frac{n_{max}}{1+e^{2A\frac{n_{max}}{2}\,(x-x_0)}} \nonumber ,
\end{align}
where we have used the identities $\tan(i\theta)=i\,\tanh(\theta)$, $1-\tanh(z)=2/1+e^{2z}$ and renamed the constant $C_2$ as $-x_0$.
Substituting \cref{eq:supp-derivation-const} into the above equation, we obtain a Fermi–Dirac-like distribution for $\rho(x)$,
\begin{equation}
\boxed{
    \rho(x)=\frac{n_{max}}{1+e^{ \frac{(x-x_{0})}{k_B T_{\rm eff}^{(\rm site)}} } }, \quad k_{B} T_{\rm eff}^{(\rm site)} = \frac{\lambda_{-}+ \lambda{+}}{2 (\lambda_{-} - \lambda_{+})},}
    \label{eq:supp-fermi-distribution-site}
\end{equation}
where $T_{\rm eff}^{(\rm site)}$ is the effective temperature for \cref{eq:supp-fermi-distribution-site}.
For clarity, we emphasize that this is distinctly different from the effective temperature $T_{eff}$ [\cref{eq:supp-fermi-distribution-curr}] obtained in \cref{sec:s1.3} below.
Note that \cref{eq:supp-fermi-distribution-site} is just the standard Fermi-Dirac distribution multiplied by $n_{max}$. To get the constant $x_0$, we use $\sum_x \rho(x) = n$ and consider the $N\to \infty$ "continuum" limit, such that
\begin{equation}
    n =\sum_{x=1}^N\rho(x)\approx\int_0^N \rho(x)dx - \int_0^1\rho(x)dx +\frac{\rho(1)+\rho(N)}{2} =\int_0^N \frac{n_{max}}{1+e^{ \frac{(x-x_{0})}{k_B T_{eff}^{(site)}} } } dx -n_{max}+\frac{n_{max}}{2},
    \label{eq:supp-particle-norm}
\end{equation}
where the term $(\rho(1) + \rho(N))/2$ accounts for the boundary correction from the Euler–Maclaurin expansion, with $\rho(1) = n_{max}$ and $\rho(N) =0$.
Solving \cref{eq:supp-particle-norm} yields
$$
n =
n_{max}\,(x_{0}-\frac{1}{2})+ n_{max}\,k_{B}T_{eff}^{(site)}\,
\ln\!\biggl[
\frac{1 + e^{-x_{0}/k_{B}T_{eff}^{(site)}}}
     {1 + e^{-(N - x_{0})/k_{B}T_{eff}^{(site)}}}
\biggr].
$$
If $N\gg k_{B}T_{eff}^{(site)}$, the logarithmic term is exponentially small (the tail beyond the step at $N=x_0$ is negligible), and one simply finds
$$
n
\;\approx\;
n_{max}(x_{0}-\frac{1}{2})
\quad\Longrightarrow\quad
\boxed{
x_{0}
\;\approx\;
\frac{n}{\,n_{max}}+\frac{1}{2}
.}
$$
In particular, for the hard‐core case $n_{max}=1$, $x_{0}\approx n+\frac{1}{2}$.

\subsection{Justification of Approximation of \cref{eq:supp-FD-approx} \label{sec:s1.2}}
We now justify the approximation [\cref{eq:supp-FD-approx}] which is
\begin{equation}
\begin{aligned}
    \lambda_{\pm}\sum_{i}\psi_{i}n_{i}(x)(n_{max}-n_{i}(x\pm1))&\approx \lambda_{\pm}\rho(x)(n_{max}-\rho(x\pm1))\notag\\
   &= \lambda_{\pm}\sum_{i}\psi_{i}n_{i}(x)\sum_{j}\psi_{j}(n_{max}-n_{j}(x\pm1)),
\end{aligned}
    \label{eq:approx justif}
\end{equation} where $\rho(x)=\sum_{i}\psi_{i}n_{i}(x)$ is the expectation value of occupation at site $x$. This is essentially the mean-field approximation where $n_i(x\pm 1)$ is approximated by $\rho(x\pm 1)=\sum_j\psi_j n_j(x\pm 1)$. 
For interacting systems, such a substitution neglects correlations between occupations at different sites.

To evaluate the validity of \cref{eq:supp-FD-approx}, we examine the left- and right-hand sides of the approximate steady-state equation \cref{eq:supp-approx-diff-eq}, which is obtained by applying the approximation \cref{eq:supp-FD-approx} to the exact steady state condition [\cref{eq:supp-interacting-HN-ss-condition}]. Since the original steady-state equation [\cref{eq:supp-interacting-HN-ss-condition}] must hold exactly, a valid approximation should yield an approximate steady-state equation \cref{eq:supp-approx-diff-eq} whose two sides remain nearly equal. The deviation between them quantifies the error introduced by our approximation [\cref{eq:supp-FD-approx}],
\begin{equation}
    \Delta_{error}=\Big(\lambda_{-}\rho(x+1)(n_{max}-\rho(x))+\lambda_{+}\rho(x-1)(n_{max}-\rho(x)) \Big)-\Big(\lambda_{-}\rho(x)(n_{max}-\rho(x-1))+\lambda_{+}\rho(x)(n_{max}-\rho(x+1))\Big).
    \label{eq:error}
\end{equation}
Since both sides of exact steady state equation [\cref{eq:supp-interacting-HN-ss-condition}] are exactly equal,
\begin{equation}
\begin{aligned}
   &\Big(\lambda_{-}\sum_{i}\psi_{i}n_{i}(x+1)(n_{max}-n_{i}(x))+\lambda_{+}\sum_{i}\psi_{i}n_{i}(x-1)(n_{max}-n_{i}(x)) \Big)-\\
    &\Big(\lambda_{-}\sum_{i}\psi_{i}n_{i}(x)(n_{max}-n_{i}(x-1))+\lambda_{+}\sum_{i}\psi_{i}n_{i}(x)(n_{max}-n_{i}(x+1)) \Big) = 0.
\end{aligned}
\end{equation}
Subtracting this equation from \cref{eq:error} on both sides, we obtain
\begin{equation}
\begin{aligned}
\Delta_{error} &=\Big(\lambda_{-}\varepsilon_{x,x+1}+\lambda_{+}\varepsilon_{x,x-1}\Big)-\Big(\lambda_{-}\varepsilon_{x,x-1}+\lambda_{+}\varepsilon_{x,x+1}\Big)\\
&= (\lambda_--\lambda_+)(\varepsilon_{x,x+1}-\varepsilon_{x,x-1}),
\end{aligned}
\label{eq:simp error}
\end{equation} where $\varepsilon_{x,x\pm1}=\sum_{i}\psi_{i}n_{i}(x\pm1)n_{i}(x)-\rho(x\pm1)\rho(x)$ is correlation measure for $n_i(x)$ and $n_i(x\pm1)$.

The approximation in \cref{eq:supp-FD-approx} becomes exact in the absence of correlations, i.e., when $\varepsilon_{x,x\pm1} = 0$, a condition generally not satisfied in interacting systems. In cases where the correlations are asymmetric ($\varepsilon_{x,x+1} \ne \varepsilon_{x,x-1}$), the resulting errors do not cancel. However, if the hopping rates are nearly symmetric, i.e., $(\lambda_{-} - \lambda_{+}) \to 0$, the asymmetry-induced error $\Delta_{error}$ remains small even in the presence of correlations. Therefore, the approximation is valid in regimes where the asymmetry between left and right hopping is weak.

\subsection{Alternative Derivation of the Fermi–like Steady State}\label{sec:s1.3}
\begin{figure}[H]
    \centering
    \includegraphics[width=0.7\linewidth]{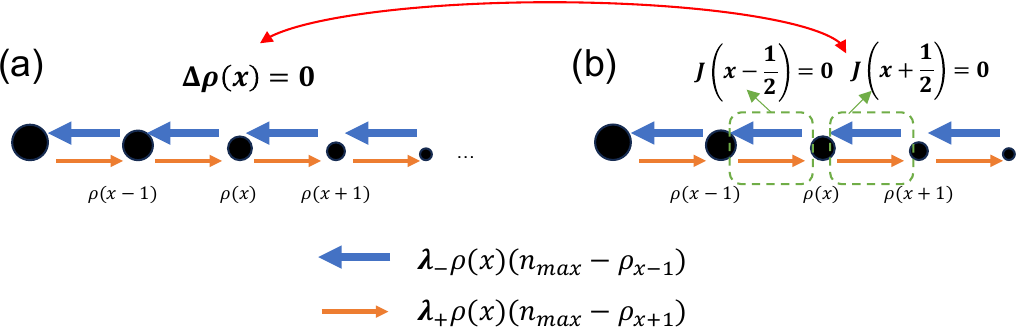}
    \caption{\tb{Equivalence of the steady-state conditions in \cref{eq:supp-approx-diff-eq} and \cref{eq:supp-ss-condition-curr}.} \cref{eq:supp-approx-diff-eq} expresses the steady-state condition from the perspective of vanishing density change at site $x$ in \tb{(a)}, while \cref{eq:supp-ss-condition-curr} formulates it in terms of zero current across the bond between $x$ and $x+1$ in \tb{(b)}.
    }
    \label{fig:supp-bond-current}
\end{figure}

Here, we offer another perspective to obtain the Fermi-like distribution.
In \cref{eq:supp-approx-diff-eq} of \cref{sec:s1.1}, the steady-state condition requires that the net change in $\rho(x)$ vanishes at each site $x$ with $\Delta \rho(x)=0$. 
We show that this condition is equivalent to having net zero current across each nearest-neighbor link (i.e., the bond connecting adjacent lattice sites) [\cref{fig:supp-bond-current}(b)]. 
Solving this improved formulation reduces truncation errors under open boundary conditions (OBC) as no Taylor expansion is required.
By rewriting \cref{eq:supp-approx-diff-eq}, the steady-state condition can be expressed as 
\begin{equation}
    \lambda_{-}\rho(x+1)(n_{max}-\rho(x))-\lambda_{+}\rho(x)(n_{max}-\rho(x+1))=\lambda_{-}\rho(x)(n_{max}-\rho(x-1))-\lambda_{+}\rho(x-1)(n_{max}-\rho(x)).
\end{equation}
We define the bond current between sites $x$ and $x+1$ as 
\begin{equation}
    J_{x+\frac{1}{2}} = \lambda_- \rho(x+1)(n_{max}-\rho(x)) - \lambda_+\rho(x)(n_{max}-\rho(x+1)).
\end{equation}
The current $J_{x+\frac{1}{2}}$ must be a constant under steady state.
Under OBC, we have $J_{1/2} = 0$, which implies
\begin{equation}
    J_{x+\frac{1}{2}} = 0.
    \label{eq:supp-ss-condition-curr}
\end{equation}
This condition leads to the recursion relation
\begin{equation}
    \frac{\rho(x+1)}{n_{max}-\rho(x+1)} = \frac{\lambda_+}{\lambda_-} \frac{\rho(x)}{n_{max}-\rho(x)},
\end{equation}
which admits the closed-form solution
\begin{equation}
\boxed{
    \rho(x) = \frac{n_{max}}{1+\exp[(x-x_0)/k_BT_{eff}^{(current)}]} \quad k_BT_{ eff}^{(current)} =\frac{1}{\ln(\lambda_-/\lambda_+)} ,}
    \label{eq:supp-fermi-distribution-curr}
\end{equation}
where $x_{0}
\;\approx\;
\frac{n}{\,n_{max}}+\frac{1}{2}$ can be similarly obtained in \cref{sec:s1.1}.

\begin{figure}[H]
    \centering
    \includegraphics[width=0.7\linewidth]{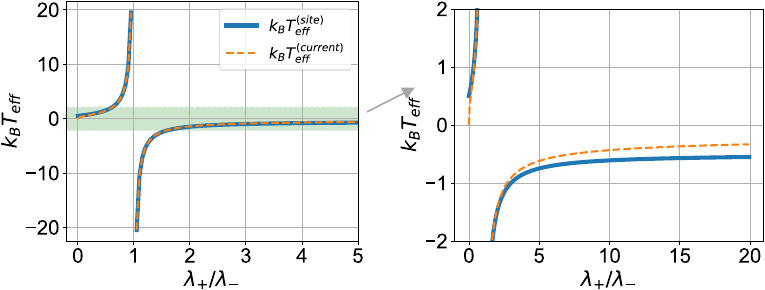}
    \caption{Asymptotic behavior of $k_B T_{eff}$ obtained from the two approaches, \cref{eq:supp-fermi-distribution-site,eq:supp-fermi-distribution-curr}. We observe that over a broad range of $\lambda_+/\lambda_-$, the two results coincide. However, in the strongly asymmetric limit, $\lambda_+/\lambda_- \to 0$ (or $\infty$), they deviate due to truncation errors introduced by the finite-order Taylor expansion.}
    \label{fig:supp-eff-temperature}
\end{figure}

Under weak Hermiticity with $\lambda_+ \approx \lambda_-$, the characteristic energy scale from the effective temperature approximate to leading order: $k_B T_{\rm eff}^{\rm current}\approx (\lambda_-+\lambda_+)/(2(\lambda_--\lambda_+))$, which matches our derivation in \cref{sec:s1.1}. 
The effective temperature in $k_B T^{\rm current}_{\text{eff}}$ depends on the ratio $\lambda_+ / \lambda_-$. 
When $\lambda_+ / \lambda_- \in (0, 1)$, the steady-state distribution resembles a positive-temperature Fermi-Dirac distribution. 
Conversely, when $\lambda_+ / \lambda_- \in (1, +\infty)$, the steady-state distribution resembles a negative-temperature Fermi-Dirac distribution. 
In the Hermitian limit with $\lambda_+ = \lambda_-$, the temperature diverges which results in a uniform distribution. 
This is verified by the near-perfect agreement between the numerical simulations and analytic results [\cref{eq:supp-fermi-distribution-curr}] in \cref{fig:supp-eff-temperature}.

\begin{figure}[H]
    \centering
    \includegraphics[width=\linewidth]{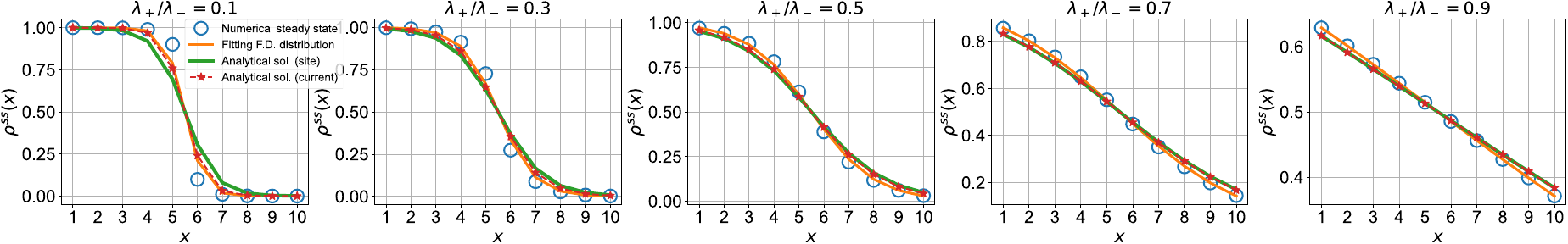}
    \caption{\textbf{Breakdown of the mean-field approximation with increasing hopping asymmetry.}
    We compare the numerically obtained steady-state distributions from time evolution [\cref{eq:supp-markov-evolve}] (blue circles) with a numerically fitted Fermi-like form (orange lines), the analytical Fermi-Dirac-like distribution given in \cref{eq:supp-fermi-distribution-site} (green lines) and \cref{eq:supp-fermi-distribution-curr} (red stars), across different values of the right hopping rate $\lambda_+$. As the hopping asymmetry (i.e., the difference between $\lambda_+$ and $\lambda_- = 1$) increases, the steady states (blue circles) increasingly deviate from the Fermi-like form (orange lines). In the weakly non-Hermitian regime, the steady states closely resemble an ideal Fermi-like distribution, and the mean-field approximation [\cref{eq:supp-FD-approx}] remains effective, as evidenced by the agreement among the numerical results (blue circles), fitted curves (orange lines), and the analytical expressions (red stars). Furthermore, the discrepancy between the two analytical results (green lines and red stars) in the strongly asymmetric regime indicates the truncation error arising from the finite-order Taylor expansion. 
    The simulations are performed with $n = 5$ particles, system size $N = 10$, and maximum occupation number $n_{max} = 1$. The panels show results for $\lambda_+ = 0.1$; $\lambda_+ = 0.3$; $\lambda_+ = 0.5$; $\lambda_+ = 0.7$ and $\lambda_+ = 0.9$. 
    }
     \label{fig:supp-FD-fitting}
\end{figure}

\clearpage
\section{Equivalence between the interacting HN model and the K-exclusion process}\label{sec:s2}
The \emph{K-exclusion process} (K-ASEP), also known as the partial exclusion process, is a generalization of the Asymmetric Simple Exclusion Process (ASEP) in which each site on a one-dimensional lattice can accommodate more than one particle, up to a finite capacity, denoted as $n_{max}$ \cite{schutz1994non,cocozza1985processus,evans2014condensation,ayyer2024exactly}. 
We consider a one-dimensional lattice with $N$ sites, where the occupation number at site $x$ is given by $n(x) \in \{0, 1, \dots, n_{max}\}$. 
The number of vacancies at site $x$ is then $n_{max} - n(x)$.
The transition probabilities governing particle hopping from site $x$ to site $x+1$ (rightward) and site $x-1$ (leftward) are given by:
\begin{equation}
P(x \to x+1) = \lambda_+ n(x) \left(n_{max} - n(x+1)\right), \
P(x \to x-1) = \lambda_- n(x) (n_{max} - n(x-1)),
\end{equation} 
where $\lambda_+ \neq \lambda_-$ introduces asymmetry in the hopping process.  
These transition probabilities ensure that a site cannot exceed its maximum occupation, as the transfer probability vanishes whenever $n(x) = 0$ or $n(x+1) = n_{max}$. 
Notably, the K-ASEP reduces to the standard ASEP when $n_{max} = 1$.

To establish the equivalence between the K-exclusion process and the interacting HN model, we begin with the master equation governing the time evolution of the probability $P_t(\mathcal{C})$ of the system being in configuration 
$\mathcal{C}$.  
Here, $\mathcal{C}$ represents an occupation configuration of particles, given by $[n(1), n(2), \dots, n(x), \dots, n(N) ]$, where $n(x)$ denotes the occupation number at site $x$. 
For a generic Markov process, the master equation describing the time evolution of $P_t(\mathcal{C})$ is 
\begin{equation}
    \frac{\mathrm{d}}{\mathrm{d}t}P_t(\mathcal{C}) = \sum_{\mathcal{C'}\neq \mathcal{C}} M(\mathcal{C}, \mathcal{C'}) P_t(\mathcal{C'})-
    \Big(\sum_{\mathcal{C'} \neq \mathcal{C}} M(\mathcal{C'},\mathcal{C}) \Big) P_t(\mathcal{C}),
\end{equation}
where $M(\mathcal{C'},\mathcal{C} )$ represents the transition probability from configuration $\mathcal{C}$ to $\mathcal{C'}$. 
Considering the specific form of the hopping rates of K-exclusion process, the transition probability from configuration $\mathcal{C}$ to $\mathcal{C'}$ is given by:
\begin{equation}
M(\mathcal{C}',\mathcal{C}) = \lambda_+ \cdot n(x) \cdot (n_{max} - n(x+1)),
\end{equation}
where the configuration spaces are $\mathcal{C}=\ket{n(1),...,n(x),n(x+1),..,n(N)} $ and $\mathcal{C'} = \ket{n(1),...,n(x)-1,n(x+1)+1,..,n(N)}$. 
The rightward and leftward particle hopping from site $x$ to site $x+1$ and $x-1$ is captured by $\lambda_+$ and $\lambda_-$, respectively.
Substituting the transition probabilities into the master equation, we obtain
\begin{equation}
    \label{eq:supp-k-exclusion}
    \begin{aligned}
        \frac{dP_t(\mathcal{C})}{dt} &= \sum_{i=1}^{L-1} \left[ \lambda_- \cdot (n(x+1)+1) \cdot (n_{max} -(n(x)-1)) P_t(\mathcal{C}^R) - \lambda_+ \cdot n(x) \cdot (n_{max} - n(x+1)) P_t(\mathcal{C}) \right] \\
        &- \lambda_- \cdot n(x) \cdot (n_{max} -n(x-1) P_t(\mathcal{C}) + \lambda_+ \cdot (n(x-1)+1) \cdot (n_{max} - (n(x)-1)) P_t(\mathcal{C}^L),
    \end{aligned}
\end{equation} 
where $\mathcal{C}=\ket{n(1), n(2),...,n(N)}$ represents the current state of the system, $\mathcal{C}^L = \ket{n(1), n(2),...,n(x-1)+1, n(x)-1,...,n(N)}$ is the state after a leftward hop, $\mathcal{C}^R = \ket{n(1), n(2),...,n(x-1), n(x)-1, n(x+1)+1,...,n(N)}$ is the state after a rightward hop.

So far, we have obtained the time evolution equation for the probability $P_t(\mathcal{C})$ of a specific configuration $\mathcal{C}$. To describe the full probability distribution over the configuration space, we enumerate all possible configurations as ${\mathcal{C}_1, \mathcal{C}_2, \dots, \mathcal{C}_\mathcal{D}}$, where $\mathcal{D}$ is the dimension of the configuration space. The probability vector is then defined as $\mathbf{P_t(\mathcal{C})} = [P_t(\mathcal{C}_1), P_t(\mathcal{C}_2),...,P_t(\mathcal{C}_\mathcal{D})]^T$, and we can rewrite \cref{eq:supp-k-exclusion} in the matrix form,
\begin{align*}
&\frac{d \mathbf{P_t(\mathcal{C})}}{dt} =-L\mathbf{P_t(\mathcal{C})} \\
&L=\sum_{x,\pm} \lambda_\pm\left(\hat \rho_x-\hat b^\dagger_{x\pm 1}\hat b_x \right)(n_{max}- \hat \rho_{x\pm 1}).
\end{align*}
Here,
$\hat b_x^\dagger|...,n(x),...\rangle = |...,n(x)+1,...\rangle$, $\hat b(x)|...,n(x),...\rangle = n(x)|...,n(x)-1,...\rangle$ and the $i$-th configuration basis is constructed as $\mathcal{C}_i = |\vec n_i\rangle = |n_i(1),n_i(2),...,n_i(N)\rangle= \Pi_{x=1}^N(\hat b_x^\dagger)^{n_i(x)}  |0\rangle$. 
A generic many-body state takes the form $\Psi(t) = \sum_{i}^\mathcal{D} P_t(\mathcal{C}_i) \mathcal{C}_i$, where $\sum_i^\mathcal{D} P(\mathcal{C}_i) = 1$ and $\mathcal{D}$ is the dimension of our configuration space. It is essentially our interacting HN model.

\section{Construction of the Laplacian for nearest-neighbor interacting spin Models}\label{sec:s3}
In this section, we detail the construction of the Laplacian operator for our interacting spin models using the Schwinger boson representation~\cite{wu1999schwinger}, which maps spin-$\frac{1}{2}$ degrees of freedom to a bosonic Fock space. 

Each spin state is encoded by a pair of occupation numbers $(n_+, n_-)$ associated with two species: the up-spin corresponds to $(1,0)$ ("+" species), and the down-spin corresponds to $(0,1)$ ("-" species).
To realize this mapping, we introduce bosonic operators ${b_{\pm}, b_{\pm}^{\dagger}}$ satisfying the canonical commutation relations $[b_{\eta}, b_{\eta'}^{\dagger}] = \delta_{\eta \eta'}$, with $\eta = \pm$. The corresponding number operators are $\hat{n}_{\pm} = b_{\pm}^{\dagger} b_{\pm}$. A many-body spin configuration over $N$ sites, $\ket{s_1, \dots, s_N}$, is thus represented in the bosonic basis as $\ket{n_{1,+}, n_{1,-}, \dots, n_{N,+}, n_{N,-}}$. Within this formalism, the spin-raising and spin-lowering operators are expressed as $s^+ = b_+^{\dagger} b_-$ and $s^- = b_-^{\dagger} b_+$, respectively, providing a bosonic realization of angular momentum operators. To obtain the Laplacian $L$ corresponding to given interactions, it suffices to include an appropriate on-site term $D$ that ensures probability conservation in the model. In the Markov formalism, probability conservation requires that each column of $L = D - H$ sums to zero, ensuring that the probability flowing out of any configuration is exactly balanced by the inflow from others.

\subsection{Non-Hermitian anti-correlated (correlated) spin-flip model} 
A general Hamiltonian describing nearest-neighbor anti-correlated and correlated spin-flip processes is given by $H=\sum_{x,\pm}t_{\pm}s_{x}^{\pm}s_{x+1}^{\pm}+\sum_{x,\pm}u_{\pm}s_{x}^{\mp}s_{x+1}^{\pm}$, where $s_x^{\pm}$ are the spin raising/lowering operators at site $x$, and $t_{\eta}, u_{\eta}$ are parameters controlling the correlated spin-flip    (pair-exchange) and spin-flip amplitudes, respectively. 
Employing the Schwinger boson representation $s_x^+ = b_{x,+}^\dagger b_{x,-},\ 
s_x^- = b_{x,-}^\dagger b_{x,+}$, we express it in bosonic form:
\begin{equation}
    H  =\sum_{x,\eta=\pm}t_{\eta}(\hat{b}_{x,\eta=\pm}^{\dagger}\hat{b}_{x,-\eta}\hat{b}_{x+1,\eta}^{\dagger}\hat{b}_{x+1,-\eta})
     +\sum_{x,\eta}u_{\eta}(\hat{b}_{x,-\eta}^{\dagger}\hat{b}_{x,\eta}\hat{b}_{x+1,\eta}^{\dagger}\hat{b}_{x+1,-\eta}),
\end{equation}

The Laplacian is constructed as $L= D-H$, where $D$ is a diagonal term to ensure probability conservation \cite{merris1994laplacian}. Specifically, we have
\begin{equation}
    \begin{aligned}
        L & =\sum_{x,\eta=\{\pm\}}t_{\eta}(\hat{n}_{x,-\eta}\hat{n}_{x+1,-\eta}-\hat{b}_{x,\eta}^{\dagger}\hat{b}_{x,-\eta}\hat{b}_{x+1,\eta}^{\dagger}\hat{b}_{x+1,-\eta})\\
        & +\sum_{x,\eta=\{\pm\}}u_{\eta}(\hat{n}_{x,\eta}\hat{n}_{x+1,-\eta}-\hat{b}_{x,-\eta}^{\dagger}\hat{b}_{x,\eta}\hat{b}_{x+1,\eta}^{\dagger}\hat{b}_{x+1,-\eta}),
    \end{aligned}
\end{equation}
where the $t_{\eta}$ term gives our non-Hermitian correlated spin-flip    model and the $u_{\eta}$ term gives the anti-correlated spin-flip   model. 

Our spin quantum Hamiltonian can be viewed as a generalized form of the celebrated quantum XY model. 
Using $s^+ = S^x+iS^y,\ s^- = S^x-iS^y$, we have
\begin{align}
    H& =\sum_{x,\pm}t_{\pm}s_{x}^{\pm}s_{x+1}^{\pm}+\sum_{x,\pm}u_{\pm}s_{x}^{\mp}s_{x+1}^{\pm} \nonumber\\
    &=\sum_x (t_+ + t_-)\big(S^x_xS_{x+1}^x - S_x^yS_{x+1}^y\big) + i (t_+-t_-)\big(S_x^xS_{x+1}^y-S_x^yS_{x+1}^x\big) + 
    \\
    & \sum_x(u_+ + u_-)\big(S^x_xS_{x+1}^x + S_x^yS_{x+1}^y\big) + i (u_+-u_-)\big(S_x^xS_{x+1}^y-S_x^yS_{x+1}^x\big). \nonumber
\end{align}
If we consider $t_{\pm}=0, u_+=u_-$, the model reduces to the Hermitian quantum XY model $H=\sum_x S_x^xS_{x+1}^x + S_x^yS_{x+1}^y$. 
When $t_+=t_-=u_+=u_-$, the Hamiltonian simplifies to the form $H = \sum_xS_x^xS_{x+1}^x$.

\subsection{Triple-spin model} Following the Schwinger boson formalism with $s^+_{\pm} = S^x_x \pm i S^y_x$, $s_x^+ = b_{x,+}^\dagger b_{x,-}$, $s_x^- = b_{x,-}^\dagger b_{x,+}$, and $S_x^z = \frac{1}{2} (\hat{b}_{x,+}^{\dagger} \hat{b}_{x,+} - \hat{b}_{x,-}^{\dagger} \hat{b}_{x,-})$,
the Laplacian of the triple spin Hamiltonian, $\hat{H} =\sum_{x}\sum_{\pm}\lambda_{\pm}(\boldsymbol{\hat{S}_{x}}\times \boldsymbol{\hat{S}_{x\pm1}})\cdot \boldsymbol{\hat{S}_{x\mp1}}$ is 
\begin{equation}
    \begin{aligned}
        L &= -\sum_x\frac{\lambda_+ - \lambda_-}{4i} \Bigg\{(n_{x,+} - n_{x,-})\cdot \Big( b_{x+1,-}^{\dagger} b_{x+1,+} b_{x-1,+}^{\dagger} b_{x-1,-} - b_{x+1,+}^{\dagger} b_{x+1,-} b_{x-1,-}^{\dagger} b_{x-1,+} \Big)  \\
        &\quad+ (n_{x+1,+} - n_{x+1,-})\cdot \Big( b_{x-1,-}^{\dagger} b_{x-1,+} b_{x,+}^{\dagger} b_{x,-} -  b_{x-1,+}^{\dagger} b_{x-1,-} b_{x,-}^{\dagger} b_{x,+} \Big)  \\
        &\quad+ (n_{x-1,+} - n_{x-1,-})\cdot \Big( b_{x,-}^{\dagger} b_{x,+} b_{x+1,+}^{\dagger} b_{x+1,-} -  b_{x,+}^{\dagger} b_{x,-} b_{x+1,-}^{\dagger} b_{x+1,+} \Big) 
        \Bigg\},
    \end{aligned}
\end{equation}
where probability conservation is enforced. 
Note the markedly different physical interpretation: while the triple spin Hamiltonian is a topological term that exists in \cite{Nickel2025,Go2025,Oh2025}, its corresponding Markov chain Laplacian contains a series of operators involving the simultaneous flipping of opposite nearby spins.

\section{Equivalence between the anti-correlated spin-flip   model and the interacting Hatano–Nelson (HN) model}\label{sec:s4}
In this section, we demonstrate that our spin-flip model $L_u$ [\cref{eq:supp-corr-spin-flip}] can be mapped to the interacting Hatano–Nelson (HN) model $L_{\lambda}$ [\cref{eq:supp-corr-interact-HN}] with $n_{max}=1$. This correspondence allows us to use known results about the interacting HN model to understand the non-equilibrium steady state of the anti-correlated spin-flip model.
\begin{align}
    L_u&=\sum_{x,\eta=\{\pm\}}u_{\eta}\Big(\hat{n}_{x,\eta}\hat{n}_{x+1,-\eta}-\hat{b}_{x,-\eta}^{\dagger}\hat{b}_{x,\eta}\hat{b}_{x+1,\eta}^{\dagger}\hat{b}_{x+1,-\eta}\Big) \label{eq:supp-corr-spin-flip} \\
    &\rotatebox{90}{$\Longleftrightarrow$} \nonumber \\ 
    L_{\lambda}&=\sum_{x}\sum_{\pm}\lambda_{\pm}\left(\hat{\rho}_{x}-\hat{b}_{x\pm1}^{\dagger}\hat{b}_{x}\right)(n_{max}-\hat{\rho}_{x\pm1}).
    \label{eq:supp-corr-interact-HN}
    \end{align} 
The operator $\hat{b}_{x,\eta}(\hat{b}_{x,\eta}^{\dagger})$ annihilates(creates) a hard-core boson of species $\eta$ at site $x$, and $\hat{n}_{x,\eta}=\hat{b}_{x,\eta}^{\dagger}\hat{b}_{x,\eta}$ is the corresponding number operator. $u_{\pm}$ are the asymmetric hopping amplitudes. Using the hard-core constraint $n_{max}=1$ to eliminate the - species boson,
$n_{x,-}=1-n_{x,+}$, 
\begin{equation}
\begin{aligned}
    L_u&= \sum_{x,\pm}u_{\pm}\left(\hat{n}_{x,+}(1-\hat{n}_{x\pm1,+})-\hat{b}_{x,+} \hat{b}_{x\pm1,+}^{\dagger} (1-\hat{n}_{x\pm1, +}) \right) \\
      &= \sum_{x}\sum_{\pm}u_{\pm}\left( \hat{n}_{x,+}-\hat{b}_{x\pm1,+}^{\dagger}\hat{b}_{x,+} \right)(1-\hat{n}_{x\pm1, +}).
\end{aligned}
\label{eq:supp-spinflip-to-interact-HN}
\end{equation}
\cref{eq:supp-spinflip-to-interact-HN}  matches the form of the interacting HN model, $L_{\lambda}$ [\cref{eq:supp-corr-interact-HN}], identifying $\hat{\rho}_x \equiv \hat{n}_{x,+}$, $\lambda_{\pm} \equiv u_{\pm}$, valid for maximum occupation number $n_{max} = 1$.  Thus, the steady-state distribution of the $+$ species boson (spin-up), denoted by $\rho^{ss}_+(x) = \langle \hat{n}_{x,+} \rangle$, should be the same as the interacting HN model, following the Fermi-Dirac distribution 
\begin{equation}
    \rho_+^{ss}(x)=\frac{1}{ 1+e^{\displaystyle\frac{2(u_{-}-u_{+} )(x-n_{tot})}{u_{-}+u_{+}}  } },
    \label{eq:supp-spinflip-ss}
\end{equation}
where $u_\pm$ serve as the right/left hopping rate of the $+$ species boson and 
$n_{\text{tot}}=\sum_x n_{x,+}$.

\begin{figure}[H]
    \centering
    \includegraphics[width=\linewidth]{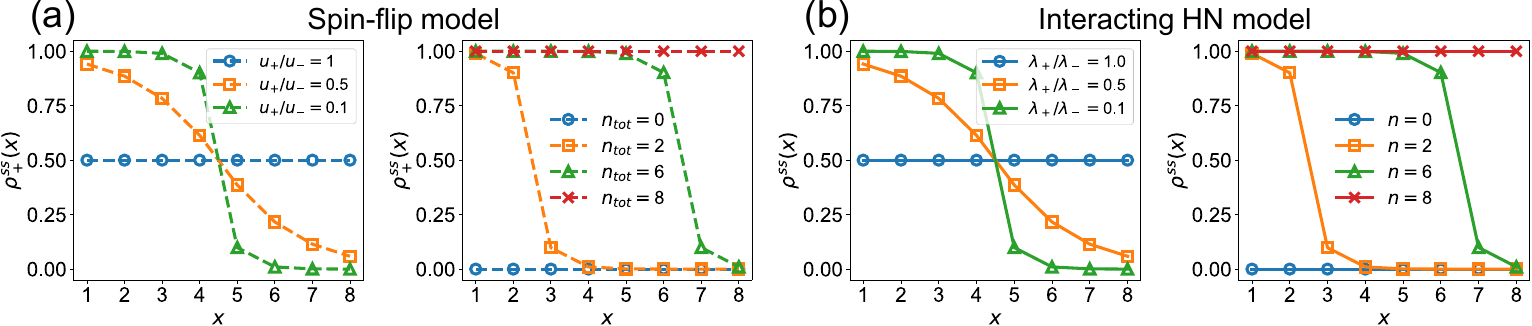}
    \caption{\textbf{Equivalence between the anti-correlated spin-flip model and the $n_{max}=1$ interacting HN model.}
    \tb{(a)} Steady-state distributions $\rho_+^{\mathrm{ss}}(x)$ [\cref{eq:supp-spinflip-ss}] for the anti-correlated spin-flip model under OBCs. Left: Results for varying hopping asymmetry ratios $u_+/u_-$ with values $1$ (blue), $0.5$ (orange), and $0.1$ (green) at fixed total number of "$+$" species bosons $n_{\mathrm{tot}} = 4$. Right: Results for varying $n_{\mathrm{tot}} = 0$ (blue), $2$ (orange), $6$ (green), and $8$ (red) at fixed $u_+/u_- = 0.1$. 
    \tb{(b)} Corresponding steady-state distributions $\rho^{\mathrm{ss}}(x)$ [\cref{eq:supp-fermi-distribution-curr}] in the interacting HN model. Left: Varying hopping asymmetries $\lambda_+/\lambda_- = 1.0$, $0.5$, and $0.1$ with fixed $n = 4$. Right: Varying particle numbers $n = 0$, $2$, $6$, and $8$ with fixed $\lambda_+/\lambda_- = 0.1$. In all panels, the system size is $N = 8$, and the maximum on-site occupation is $n_{max} = 1$. }
    \label{fig:supp-spin-flip-ss}
\end{figure}

\section{Derivation of staggered occupation in the steady state for the correlated spin-flip model with an even number of sites $N$}\label{sec:s5}

\subsection{Review of the Model and Main Results}\label{sec:s5.1}
\paragraph{Model introduction} The Laplacian of the correlated spin-flip model is given by
\begin{equation}
    L=\sum_{x,\eta=\{\pm\}}t_{\eta}(\hat{n}_{x,-\eta}\hat{n}_{x+1,-\eta}-\hat{b}_{x,\eta}^{\dagger}\hat{b}_{x,-\eta}\hat{b}_{x+1,\eta}^{\dagger}\hat{b}_{x+1,-\eta}),
    \label{eq:supp-pair-flip-Lap}
\end{equation}
where $\hat{b}_{x,\eta}$/$\hat{b}_{x,\eta}^{\dagger}$ annihilates/creates a boson of species $\eta=\{\pm\}$ at site $x$, and $\hat{n}_{x,\eta} = \hat{b}_{x,\eta}^{\dagger} \hat{b}_{x,\eta}$ is the corresponding number operator. Non-Hermiticity is introduced when $t_+\neq t_-$, and we define $\gamma:=t_+/t_-$ for later convenience.
We consider a 1D chain of length $N$ with hard-core constraint: each site hosts exactly one boson, enforcing $n_{x,+} + n_{x,-} = 1$. For notational convenience, we work in the occupation number basis of the $+$ species, denoted by $\ket{n_{1,+}, n_{2,+}, \dots, n_{N,+}}$. Since $n_{x,+} \in \{0,1\}$ under the hard-core condition, basis states correspond to binary strings, e.g., $\ket{0,1,1,0,\dots}$.

\begin{figure}[H]
    \centering
    \includegraphics[width=0.8\linewidth]{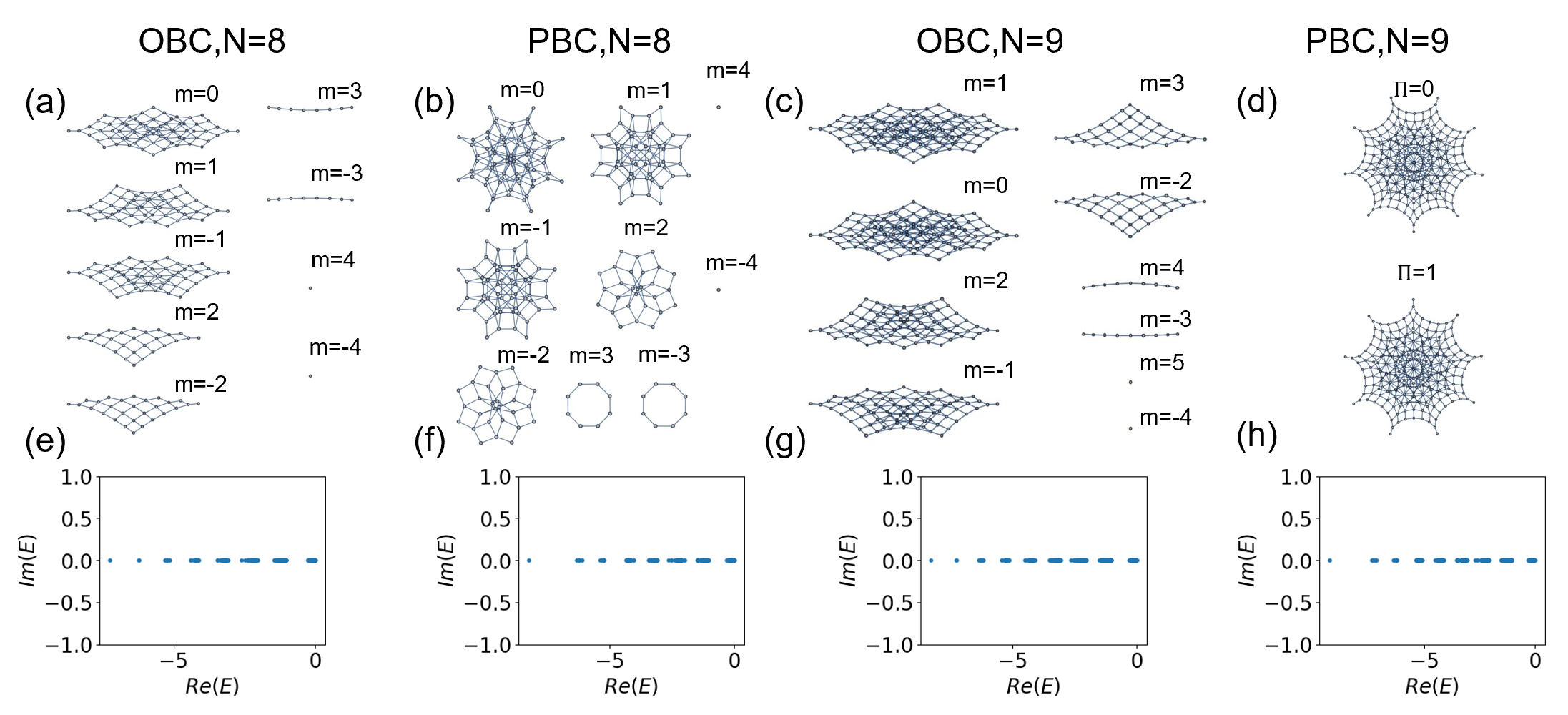}
    \caption{State-space fragmentation as shown in state connectivity graphs and the eigenvalue spectra of $-L$ [\cref{eq:supp-pair-flip-Lap}] for systems of even ($N=8$) and odd sizes ($N=9$) under periodic (PBCs) and open boundary conditions (OBCs), where right and left hopping rates are $t_{+}=0.1,t_{-}=1$. Different $m$ sectors correspond to disconnected components of the state space, except under PBC with odd $N$.
    For this exception, the disconnected subspaces are labeled by $\Pi$, where $\Pi=\{0,1\}$ is defined as \cref{eq:supp-boson-number-parity} in \cref{sec:s6.1} and denotes the parity of the number of $+$ bosons. 
    }
    \label{fig:supp-pair-flip-graph}
\end{figure}

\paragraph{State space fragmentation and staggered occupation in steady states}
For even-sized systems, the pair-flipping term $\hat{b}_{x,\eta}^{\dagger}\hat{b}_{x,-\eta}\hat{b}_{x+1,\eta}^{\dagger}\hat{b}_{x+1,-\eta}$ acts as a local kinetic constraint, allowing spin exchange only in pairs at neighboring sites. 
This leads to a fragmentation of the many-body state space into $N+1$ dynamically disconnected sectors, as shown in \cref{fig:supp-pair-flip-graph}(a-b), 
characterized by the dynamic invariant \emph{even-odd imbalance} \cite{michael2015observation}, which is equivalent to the antiferromagnetic order in quantum spin systems:
\begin{equation}
    m =\sum_{x=1}^{N} (-1)^{x+1} n_{x,+} =\sum_{x=odd}n_{x,+}-\sum_{x=even}n_{x,+}.
\label{eq:supp-even-odd-imbalance}
\end{equation} 

This state-space fragmentation yields $N+1$ distinct steady states [\cref{fig:supp-pair-flip-even-size-ss}(a)], each labeled by $m$. It is reminiscent of classical kinetically constrained models such as the Fredrickson–Andersen model, where local constraints similarly fragment the state space and give rise to non-ergodic behavior \cite{fredrickson1984kinetic}. 
The steady-state profile of the $+$ species boson's occupation distribution, 
\begin{equation}
    \rho_+^{ss}(x) = \langle n_{x,+}\rangle,
\end{equation}
exhibit a staggered pattern where
\begin{equation}
\rho_+^{ss}(x)=
\begin{cases}
   \rho_+^{(o)} & \text{if }x \text{ is odd} \\
   \rho_+^{(e)} & \text{if }x \text{ is even.} 
\end{cases}
\end{equation}
Below in \cref{sec:s5.2}, we derive their analytical expressions
\begin{equation}
\begin{aligned}
    \rho_+^{(e)}&=\frac{\sum_{i=0}^{l-m}\binom{l}{i+m}\binom{l}{i} \gamma^{i} \cdot (\frac{i}{l}) }{\sum_{i=0}^{l-m}\binom{l}{i+m}\binom{l}{i} \gamma^{i}};\\
    \rho_+^{(o)} &=\frac{\sum_{i=0}^{l-m}\binom{l}{i+m}\binom{l}{i} \gamma^{i} \cdot (\frac{i+m}{l})}{\sum_{i=0}^{l-m}\binom{l}{i+m}\binom{l}{i} \gamma^{i}}=\rho_+^{(e)}+\frac{2m}{N},
    \end{aligned}
\label{eq:supp-pair-flip-model-ss} 
\end{equation}
with $\Delta \rho_+^{ss} = \rho_+^{(o)}-\rho_+^{(e)} = 2m/N$, where $l=\left \lfloor N/2 \right \rfloor$ is half the system size and $\gamma:=t_+/t_-$ is the asymmetric hopping ratio.

\subsection{Derivation of the staggered occupation profile in the steady state for even $N$-sized systems}\label{sec:s5.2}
In this section, we derive the analytic form of the steady state [\cref{eq:supp-pair-flip-model-ss}] for the correlated spin-flip    model,  
\[
 L=\sum_{x,\eta=\{\pm\}}t_{\eta}(\hat{n}_{x,-\eta}\hat{n}_{x+1,-\eta}-\hat{b}_{x,\eta}^{\dagger}\hat{b}_{x,-\eta}\hat{b}_{x+1,\eta}^{\dagger}\hat{b}_{x+1,-\eta}),
\]
for systems with an even number of sites. We define $\gamma:=t_+/t_-$ as the asymmetric hopping ratio and $l= \left \lfloor N/2 \right \rfloor$ as half the system size for convenience.

\begin{figure}[htbp]
    \centering
    \includegraphics[width=0.9\textwidth]{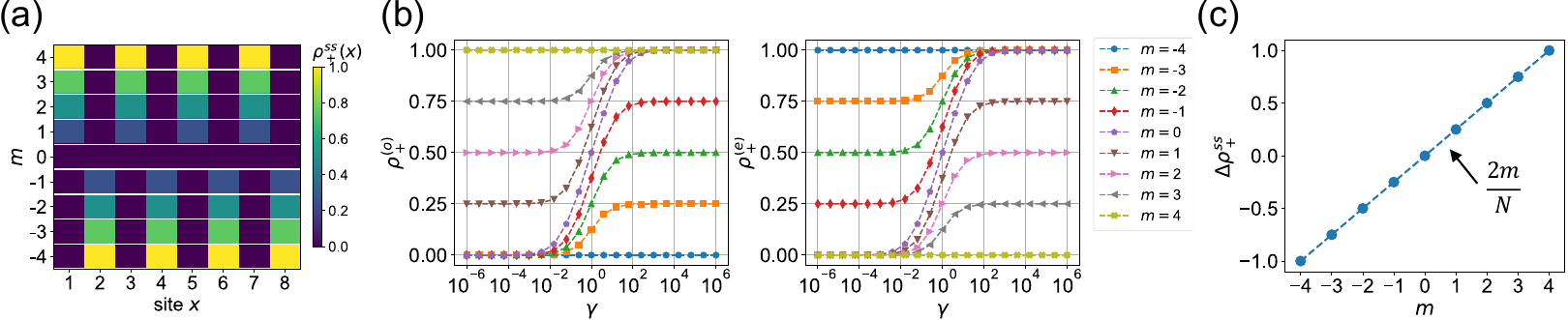}
    \caption{\textbf{Staggered occupation in the steady states of even-sized systems and their dependence on the asymmetry hopping ratio $\gamma = t_+/t_-$ in the correlated spin-flip    model.} 
    \tb{(a)} The steady-state occupation density $\rho_+(x)=\langle n_{x,+}\rangle$ exhibits a characteristic staggered occupation pattern for a system of size $N=8$ under OBCs and PBCs. 
    \tb{(b)} The asymptotic behavior of the steady states [\cref{eq:supp-pair-flip-ss-even}] on odd sites (left, $\rho_+^{(o)}$) and even sites (right, $\rho_+^{(e)}$)  is shown as a function of the asymmetry ratio $\gamma = t_+/t_-$ and also summarized in \cref{tab:supp-rho-delta-limits-even}. Tuning $\gamma$ from very small to very large only shifts both $\rho_+^{(e)}$ and $\rho_+^{(o)}$ by $2/N=0.25$. 
    \tb{(c)} The amplitude of the spatial fluctuations of the steady states, defined as $\Delta \rho_+^{ss}=\rho_+^{(o)} - \rho_+^{(e)}$, follows a linear relationship with the dynamic invariant $m$ [\cref{eq:supp-even-odd-imbalance}], given by $\Delta \rho_+^{ss} =2m/N$. 
    The asymmetry hopping ratio for \tb(a) is set as $\gamma=10^{-6}$.  
    }
\label{fig:supp-pair-flip-even-size-ss}
\end{figure}

\paragraph{Dimension of each disconnected sector.}
The correlated spin-flip term in the Laplacian imposes a local dynamical constraint, thereby partitioning the full state space into dynamically disconnected sectors, each labeled by a distinct $m$.
As introduced in \cref{eq:supp-even-odd-imbalance}, we define a dynamic invariant $m = A - B$, where $A$ and $B$ denote the number of occupied sites at odd and even indices, within a given binary basis state.
Since $A$ and $B$ each ranges from 0 to $l$, the allowed values of $m = A - B$ ranges over $\{-l, -l+1, \dots, l\}$, resulting in $2l + 1$ disjoint sectors [\cref{fig:supp-pair-flip-graph}(a,b)]. Here, $l = \left\lfloor N/2 \right\rfloor=N/2$ denotes half the system size.
The dimension of each $m$-sector, denoted by $\mathcal{D}_m$(i.e., the number of nodes in one subgraph associated with $m$ [\cref{fig:supp-pair-flip-graph}(a,b)]) is given by 
\begin{equation}
    \mathcal{D}_m = \sum_{A=0}^{l} \binom{l}{A} \binom{l}{B} = \sum_{A=0}^{l} \binom{l}{A} \binom{l}{A-m} = \binom{2l}{l+m},
\label{eq:supp-dim-m}
\end{equation} 
which counts the number of basis states with a constraint of $A - B = m$. Summing over all $m$, we recover the total state space dimension, $\sum_m \mathcal{D}_m = \sum_{m=-l}^{l} \binom{2l}{l + m} = 2^{2l}$. 

\begin{figure}[htbp]
 \centering
 \includegraphics[width=0.9\linewidth]{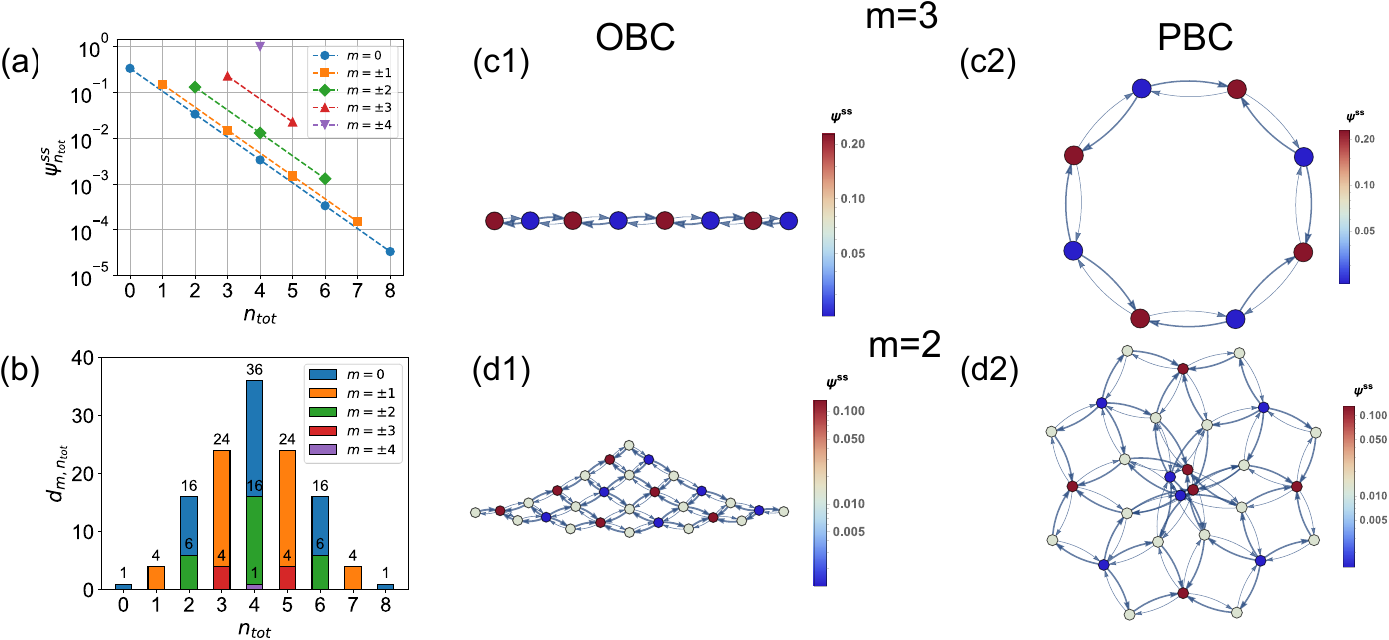}
 \caption{\tb{Distribution of the steady state $\psi^{ss}$ in state space.}
 \tb{(a)} The steady-state wavefunction $\psi^{\mathrm{ss}}_{n_{\rm tot}}$ exhibits exponential localization with respect to $n_{\rm tot}$ [\cref{eq:supp-pair-flip-ss-exp}].
 \tb{(b)} Degeneracy $d_{m,n_{tot}}$ defined in \cref{eq:supp-degeneracy}.
 \tb{(c-d)} Schematic illustration of steady states $\psi^{ss}$ on the state-space connecting graph for $m=3$ and $m=2$, and the arrangement of non-reciprocal hoppings under OBC (left panels \textbf{(c1) (d1)}) and PBC (right panels \textbf{(c2), (d2)}). Evidently, the boundary conditions significantly affect the shape of the state space graph. 
 The asymmetric hopping ratio is $\gamma=t_+/t_-=0.1$ and the system size is $N=8$.}
 \label{fig:supp-pair-flip-ss-graph}
\end{figure}

\paragraph{Exponential decay of the steady state wavefunction.} 
We now further examine the structure and profile of the steady-state wavefunction. Within each $m$-sector, we can further classify basis states by the total occupation number 
\begin{equation}
    n_{\rm tot}=A+B,
\end{equation}
where $A$ and $B$ represent the number of occupied sites at odd and even positions, respectively. The pair-flipping dynamics [\cref{eq:supp-pair-flip-Lap}] constrain $n_{\rm tot}$ to vary in steps of 2 within the range
\begin{equation}
    n_{tot}=|m|, |m|+2,...,2l-|m|,
 \label{eq:allowed n_{tot}}
\end{equation} 
where $l = \lfloor N/2\rfloor$ is the number of odd (or even) sites.
For basis states associated with given $n_{tot}$, the number of occupied sites at odd and even positions are $A = \frac{n_{tot}+m}{2},\ B = \frac{n_{tot}-m}{2}$. 
The number of basis states with fixed total occupation $n_{tot}$ in the $m$-sector, denoted by $d_{m,n_{tot}}$, is therefore 
\begin{equation}
    d_{m,n_{tot}}=\binom{l}{A}\binom{l}{B}=\binom{l}{\frac{n_{tot}+m}{2}}\binom{l}{\frac{n_{tot}-m}{2}},
\label{eq:supp-degeneracy}
\end{equation}
as shown in \cref{fig:supp-pair-flip-ss-graph}(b), given further interpretation to \cref{eq:supp-dim-m}. It counts all ways to distribute $A$ and $B$ occupied sites in $l$ odd-indexed and $l$ even-indexed positions, respectively. The dynamics within each $m$-sector are in general ergodic, and by the Perron–Frobenius theorem, each subspace admits a unique steady state associated with a zero eigenvalue of $-L$, while all other eigenvalues have negative real parts \cite{bhatia2013matrix}.
Assuming that detailed balance holds in the steady state within each $m$-sector, we consider transitions between basis states that differ by a single pair-flipping process. For example, $\ket{\dots, 0, 0, \dots} \rightarrow \ket{\dots, 1, 1, \dots}$ increases the total occupation number $n_{tot}$ by 2 while preserving the invariant $m$. The steady-state wavefunction should satisfy $\frac{\psi^{ss}_{n_{tot}+2}}{\psi^{ss}_{n_{tot}}} = \frac{t_+}{t_-}=\gamma$, where $\psi^{ss}_{n_{tot}}$ is the probability amplitude for basis states with $n_{tot}$ and $\gamma$ is the asymmetric hopping ratio.
Consequently, if we consider all possible $n_{tot}$ [\cref{eq:allowed n_{tot}}] in $m$-sector, the steady-state wavefunction follows an exponential pattern [\cref{fig:supp-pair-flip-ss-graph} (a)]: 
\begin{equation} 
    \left[\psi_{n_{tot}=|m|}, \psi_{n_{tot}=|m|+2}, ...,\psi_{n_{tot}=2l-|m|}\right] = \frac{1}{\mathcal{N}_m}\left[1, \gamma, \dots, \gamma^{l-|m|}\right],
    \label{eq:supp-pair-flip-ss-exp}
\end{equation} 
where $\mathcal{N}_m$ is the normalization factor. Note that the degeneracy of the state $\psi_{n_{tot}}$ is $d_{m,n_{tot}}$ [\cref{eq:supp-degeneracy}] in the $m$-sector. Probability normalization requires 
\begin{equation}
    \sum_{n_{\rm tot}} \psi_{n_{\rm tot}} d_{m,n_{\rm tot}} = 1,
\end{equation}
yielding the normalization constant 
\begin{equation}
     \mathcal{N}_m = \sum_{i=0}^{l-m}\binom{l}{i+m}\binom{l}{i} \gamma^{i},
    \label{eq:even prob norm}
\end{equation} 
where we rewrite the degeneracy of states as $d_{m,n_{tot}}=\binom{l}{i+m}\binom{l}{i}$ [\cref{eq:supp-degeneracy}] with $i = (n_{\rm tot}-m)/2$.

\paragraph{Staggered occupation pattern in the steady state.} 
For convenience, we consider $m>0$, as the case with  $m\le 0$ follows by switching odd and even sites. A basis state labeled by $m$ with $n_{tot}=m+2i$
contains $A=m+i$ particles on odd sites and $B=i$ particles on even sites. Since each sublattice (odd/even sites) contains $l$ sites, the average occupation per odd site in this basis state is $\frac{i+m}{l}$. Taking into account the degeneracy $d_{m,n_{tot}}=\binom{l}{i+m}\binom{l}{i}$ [\cref{eq:supp-degeneracy}] and the steady-state probability weight $\gamma^i \frac{1}{\mathcal{N}_m}$, the expected occupation per odd site in the steady state is $\sum_i \binom{l}{i+m}\binom{l}{i}(\gamma)^i \frac{1}{\mathcal{N}_m} \cdot \frac{i+m}{l}$. Similarly, the expected occupation at per even site is $\sum_i\binom{l}{i+m}\binom{l}{i}(\gamma)^i \frac{1}{\mathcal{N}_m} \cdot \frac{i}{l}$. 
Substituting $\frac{1}{\mathcal{N}_m}$ from \cref{eq:even prob norm}, we obtain the piecewise expression for the steady-state occupation profile $\rho_+^{ss}(x)= \langle n_{x,+}\rangle$ as follows:
\begin{equation}
\rho_+^{ss}(x)=
\begin{cases}
   \rho_+^{(o)} & \text{if }x \text{ is odd,} \\
   \rho_+^{(e)} & \text{if }x \text{ is even,} 
\end{cases}
\end{equation}
with
\begin{equation}
\begin{cases}
    \rho_+^{(o)} = \sum_{i=0}^{l-m}\binom{l}{i+m}\binom{l}{i} \gamma^{i} (\frac{i+m}{l}) /\mathcal{N}_{m} \\
    \rho_+^{(e)} = \sum_{i=0}^{l-m}\binom{l}{i+m}\binom{l}{i} \gamma^{i} (\frac{i}{l}) /\mathcal{N}_{m}
\end{cases} \text{\ if} \ m>0,
\end{equation}
and
\begin{equation}
\begin{cases}
    \rho_+^{(o)} = \sum_{i=0}^{l-|m|}\binom{l}{i+|m|}\binom{l}{i} \gamma^{i} (\frac{i}{l}) /\mathcal{N}_{m} \\
    \rho_+^{(e)} = \sum_{i=0}^{l-|m|}\binom{l}{i+|m|}\binom{l}{i} \gamma^{i} (\frac{i+|m|}{l}) /\mathcal{N}_{m}
\end{cases} \text{\ if} \ m\leq0,
\end{equation}
where $\mathcal{N}_{m} = \sum_{i=0}^{l-|m|}\binom{l}{i+|m|}\binom{l}{i} \gamma^{i}$.

As $\binom{l}{i+m}=0$ when $i+m>l$, we can unify above expressions for even-sized systems as
\begin{equation}
\rho_+^{ss}(x)=
\begin{cases}
   \rho_+^{(o)} & \text{if }x \text{ is odd} \\
   \rho_+^{(e)} & \text{if }x \text{ is even,} 
\end{cases}
\end{equation}
such that
\begin{equation}
\begin{aligned}
    \rho_+^{(o)} &=\frac{\sum_{i=0}^{l-m}\binom{l}{i+m}\binom{l}{i} \gamma^{i} \cdot (\frac{i+m}{l})}{\sum_{i=0}^{l-m}\binom{l}{i+m}\binom{l}{i} \gamma^{i}}; \\
    \rho_+^{(e)}&=\frac{\sum_{i=0}^{l-m}\binom{l}{i+m}\binom{l}{i} \gamma^{i} \cdot (\frac{i}{l}) }{\sum_{i=0}^{l-m}\binom{l}{i+m}\binom{l}{i} \gamma^{i}}, 
    \label{eq:supp-pair-flip-ss-even}
\end{aligned}
\end{equation}
where $l=\left \lfloor N/2 \right \rfloor$ is half the system size and $\gamma:=t_+/t_-$ is the asymmetric hopping ratio. Additionally, the steady-state amplitude, defined as  $\Delta \rho_+^{ss} = \rho_+^{(o)} - \rho_+^{(e)}$, 
exhibits a linear relationship with the dynamic invariant $m$, such that  
\begin{equation}
    \Delta \rho_+^{ss} = \frac{m}{l}.
\end{equation}

We briefly explain why the steady states remain unchanged under a change from OBC to PBC [\cref{fig:supp-pair-flip-even-size-ss}]. Under OBC, the directed transitions within each subgraph are locally arranged in opposing orientations, shown in \cref{fig:supp-pair-flip-ss-graph} (c1,d1). 
When switching to PBC, this destructive arrangement (in terms of non-reciprocal hopping direction) prevents the formation of closed directed loops, as illustrated in \cref{fig:supp-pair-flip-ss-graph} (c2,d2), such that no NHSE accumulation can take place.  
As a result, no net current is induced under PBC, and the steady state remains unchanged.
Additionally, the spectra of $L$ under PBC remain purely real, indicating the absence of oscillatory dynamics [\cref{fig:supp-pair-flip-graph}].

\subsection{Asymptotic behavior of the staggered occupation in the steady state with an even number of sites $N$}\label{sec:s5.3}
We now analyze the asymptotic behavior of the steady-state occupation at even sites, $\rho_+^{(e)}$ [\cref{eq:supp-pair-flip-ss-even}], in three regimes. The occupation on odd sites is related via $\rho_+^{(o)} = \rho_+^{(e)} + \frac{m}{l}$, where $l=\left \lfloor N/2 \right\rfloor$. We start with $m>0$ first.

\tb{1. $\gamma\to0$:}

When $\gamma\to0$, the $i=0$ term dominates both sums.  Expanding $\rho_+^{(e)}$ to the first order in $\gamma$, we have
\begin{equation}
    \rho_+^{(e)} = \frac{\sum_{i=0}^{l-m}\mathcal{F}_i \gamma^i \frac{i}{l}}{\sum_{i=0}^{l-m}\mathcal{F}_i\,\gamma^i}=\frac{\mathcal{F}_1\,(1/l)\,\gamma}{\mathcal{F}_0 + O(\gamma)}
=\frac{l\,(l-m)}{m+1}\,\frac{1}{l}\,\gamma
+O(\gamma^2)
=\frac{l-m}{m+1}\,\gamma
+O(\gamma^2)
\xrightarrow[\gamma\to 0]{}  0,
\end{equation}
where $\mathcal{F}_i=\binom{l}{i+m}\binom{l}{i}$.

\tb{2. $\gamma=1$:}

For $\gamma=1$, the weights $\mathcal{F}_i$ are symmetric under $i\mapsto(l-m)-i$, so

$$
\sum_{i}i\,\mathcal{F}_i
=\frac{l-m}{2}\,\sum_i \mathcal{F}_i,
$$
and therefore
\begin{equation}
\rho_+^{(e)}\big|_{\gamma=1}
=\frac{1}{l}\,\frac{\sum_i i\,\mathcal{F}_i}{\sum_i \mathcal{F}_i}
=\frac{1}{l}\,\frac{\tfrac{l-m}{2}\sum_i \mathcal{F}_i}{\sum_i \mathcal{F}_i}
=\frac{l-m}{2l}.
\end{equation}

\tb{3. $\gamma\to\infty$:}

For $\gamma\gg1$,
\begin{equation}
    \rho_+^{(e)} = \frac{\sum_{i=0}^{l-m}\mathcal{F}_i \gamma^i \frac{i}{l}}{\sum_{i=0}^{l-m}\mathcal{F}_i\,\gamma^i}
\xrightarrow[\gamma\to \infty]{}  \frac{\mathcal{F}_{l-m}\frac{l-m}{l}}{\mathcal{F}_{l-m}} = \frac{l-m}{l}.
\end{equation}
The case for $m\leq 0$ is similar. Given $\Delta\rho^{ss}_+ = \rho_+^{(o)}-\rho^{(e)}_+=\frac{m}{l}$, the asymptotic behavior of the steady states for even system sizes is summarized as follows:\\

\renewcommand{\arraystretch}{2}
\setlength{\tabcolsep}{10pt}
\begin{table}[htbp]
\centering
\begin{tabular}{c c c c c}
\toprule
Limit & Case & $\rho_+^{(o)}$ & $\rho_+^{(e)}$ & $\Delta \rho_+^{ss}=\rho_+^{(o)}-\rho_+^{(e)}$ \\
\midrule
$\gamma \to 1$      &    & $\displaystyle \frac{l+m}{2l}$            & $\displaystyle\frac{l-m}{2l}$                               & $\displaystyle \frac{m}{l}$          \\
$\gamma \to 0$      & $m>0$   & $\displaystyle \frac{m}{l}$            & $0$                               & $\displaystyle \frac{m}{l}$          \\
$\gamma \to 0$      & $m\le0$ & $0$                                     & $\displaystyle \frac{-m}{l}$      & $\displaystyle \frac{m}{l}$         \\
$\gamma \to \infty$ & $m>0$   & $1$                                     & $\displaystyle \frac{l-m}{l}$    & $\displaystyle \frac{m}{l}$           \\
$\gamma \to \infty$ & $m\le0$ & $\displaystyle \frac{l+m}{l}$      & $1$                               & $\displaystyle \frac{m}{l}$     \\
\bottomrule
\end{tabular}
\caption{Asymptotic steady-state occupations at odd ($\rho_+^{(o)}$) and even sites ($\rho_+^{(o)}$) for even $N$-sized systems. $l=\left\lfloor N/2 \right\rfloor$ is the half system size and $\gamma:=t_+/t_-$ is the asymmetric hopping ratio.
}
\label{tab:supp-rho-delta-limits-even}
\end{table}
\newpage

\section{Derivation of the steady states for the correlated spin-flip model with an odd number of sites $N$}\label{sec:s6}
\subsection{Review of the Model and Main Results}\label{sec:s6.1}
\paragraph{Model introduction} 
This section is about the same model as described in the previous section (\cref{sec:s5}), but we repeat the relevant definitions and notation again below for convenience for the reader.
The Laplacian of the correlated spin-flip    model is given by
\begin{equation}
    L=\sum_{x,\eta=\{\pm\}}t_{\eta}(\hat{n}_{x,-\eta}\hat{n}_{x+1,-\eta}-\hat{b}_{x,\eta}^{\dagger}\hat{b}_{x,-\eta}\hat{b}_{x+1,\eta}^{\dagger}\hat{b}_{x+1,-\eta}),
    \label{eq:supp-pair-flip-Lap}
\end{equation}
where $\hat{b}_{x,\eta}$/$\hat{b}_{x,\eta}^{\dagger}$ annihilates/creates a boson of species $\eta=\{\pm\}$ at site $x$, and $\hat{n}_{x,\eta} = \hat{b}_{x,\eta}^{\dagger} \hat{b}_{x,\eta}$ is the corresponding number operator. Non-Hermiticity is introduced when $t_+\neq t_-$, and we define $\gamma:=t_+/t_-$ for later convenience.
We consider a 1D chain of length $N$ with a hard-core constraint: each site hosts exactly one boson, enforcing $n_{x,+} + n_{x,-} = 1$. For notational convenience, we work in the occupation number basis of the $+$ species, denoted by $\ket{n_{1,+}, n_{2,+}, \dots, n_{N,+}}$. Since $n_{x,+} \in \{0,1\}$ under the hard-core condition, basis states correspond to binary strings, e.g., $\ket{0,1,1,0,\dots}$. A generic many-body state is given by $\Psi=\sum_{\vec n}\psi_{\vec n} |\vec n\rangle$ where $|\vec n\rangle = |n_{1,+},...,n_{N,+}\rangle$.

\paragraph{Staggered occupation in steady state under OBCs}
In contrast to even-sized systems, odd-sized systems necessitate separate analyses under open (OBCs) and periodic boundary conditions (PBCs). Under OBCs, the \emph{even-odd imbalance} defined in \cref{eq:even-odd-imbalance},
\begin{equation}
    m=\sum_{x=1}^{N} (-1)^{x+1} n_{x,+}
\end{equation}
remains a dynamical invariant. Following similar analysis in \cref{sec:s5.1,sec:s5.2}, the state space fragments into $N+1$ disjoint sectors labeled by $m$. The steady-state occupation profile, $\rho_+^{ss}(x)= \langle n_{x,+}\rangle$, exhibits a staggered pattern under OBCs,
\begin{equation}
\rho_+^{ss}(x)=
\begin{cases}
   \rho_+^{(o)} & \text{if }x \text{ is odd,} \\
   \rho_+^{(e)} & \text{if }x \text{ is even,} 
\end{cases}
\label{eq:supp-pair-flip-ss-odd-obc}
\end{equation}
with $\rho_+^{(o)}$ and $\rho_+^{(e)}$ determined by $m$ as follows:
\begin{equation}
\begin{cases}
    \rho_+^{(o)} = \sum_{i=0}^{l+1-m}\binom{l+1}{i+m}\binom{l}{i} \gamma^{i} (\frac{i+m}{l+1}) /\mathcal{N}_{m}^+ \\
    \rho_+^{(e)} = \sum_{i=0}^{l+1-m}\binom{l+1}{i+m}\binom{l}{i} \gamma^{i} (\frac{i}{l}) /\mathcal{N}_{m}^+
\end{cases} \text{ if } \ m>0,
\label{eq:supp-pair-flip-ss-odd-obc-m>0}
\end{equation} 
where $\mathcal{N}_{m}^+ = \sum_{i=0}^{l+1-m}\binom{l+1}{i+m}\binom{l}{i} \gamma^{i}$,
\begin{equation}
\begin{cases}
    \rho_+^{(o)} = \sum_{i=0}^{l-|m|}\binom{l+1}{i}\binom{l}{i+|m|} \gamma^{i} (\frac{i}{l+1}) /\mathcal{N}_{m}^- \\
    \rho_+^{(e)} = \sum_{i=0}^{l-|m|}\binom{l+1}{i}\binom{l}{i+|m|} \gamma^{i} (\frac{i+|m|}{l}) /\mathcal{N}_{m}^-
\end{cases} \text{ if } \ m\leq0,
\label{eq:supp-pair-flip-ss-odd-obc-m<0}
\end{equation}
where $\mathcal{N}_{m}^- = \sum_{i=0}^{l-|m|}\binom{l+1}{i}\binom{l}{i+|m|} \gamma^{i}$, $\gamma:=t_+/t_-$ and $l=\left\lfloor N/2 \right \rfloor$ is half the system size.
Compared with \cref{eq:supp-pair-flip-ss-even} in \cref{sec:s5.2}, the odd-sized systems break the sublattice-exchange symmetry. For even $N$, the transformation $o\leftrightarrow e$ together with $m\mapsto -m$ leaves the steady state invariant and implies $\rho_{+}^{(o)}(m)=\rho_{+}^{(e)}(-m)$ and $\rho_{+}^{(e)}(m)=\rho_{+}^{(o)}(-m)$ [\cref{eq:supp-pair-flip-ss-even}]. For odd $N$, these equalities no longer hold because the odd and even sublattices have unequal sizes ($l+1$ vs.\ $l$), leading to different normalizations and summation ranges.

\begin{figure}[htbp]
    \centering
    \includegraphics[width=0.9\textwidth]{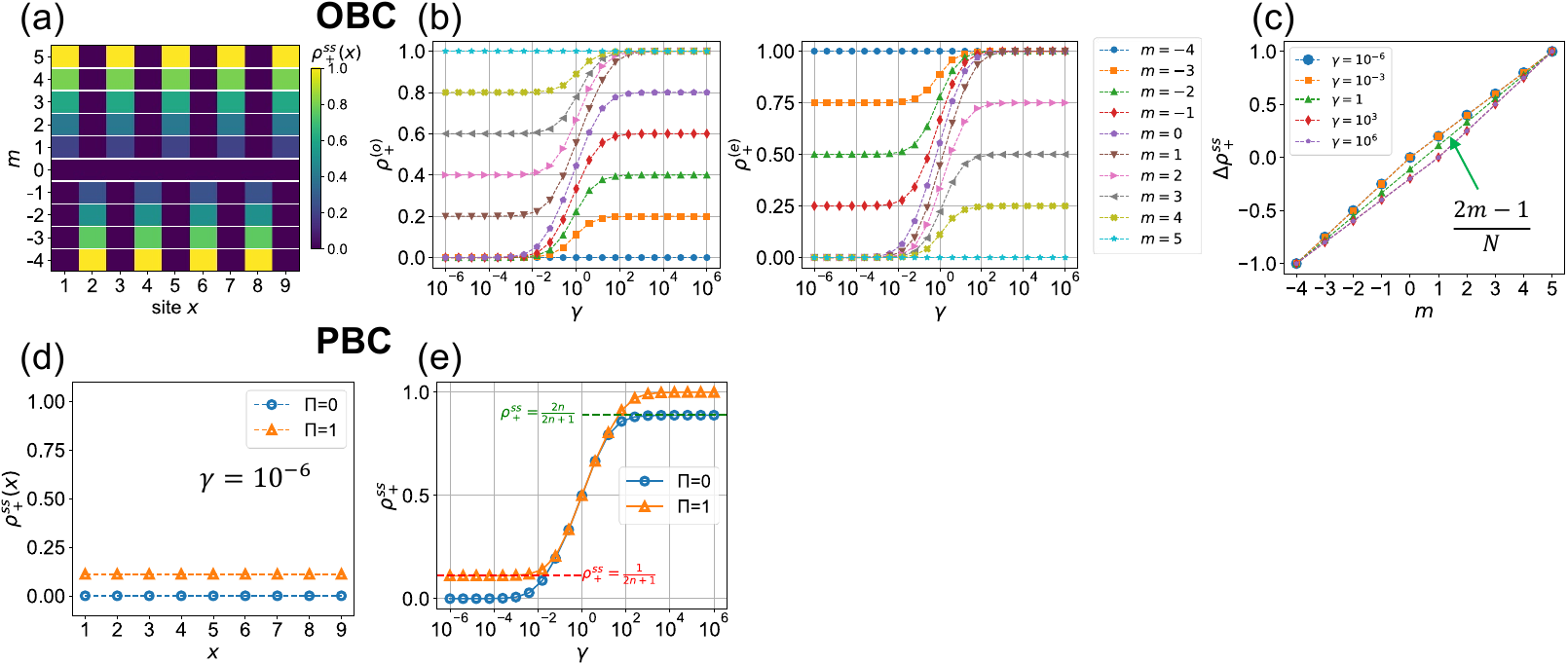}
    \caption{\tb{Comparison of the steady states $\rho_+^{ss}(x)$ [\cref{eq:supp-pair-flip-ss-odd-obc}] under OBC and PBC for systems with an odd number of sites $N$.} 
    \textbf{(a)} Under OBC, the steady states with different $m$ exhibit the staggered occupation pattern. 
    \textbf{(b)} Asymptotic behavior of the steady-state densities at odd ($\rho_+^{(o)}$, left) and even ($\rho_+^{(e)}$, right) sites for various values of $\gamma = t_+/t_-$. The asymptotic behavior of the steady states at $\gamma \to \infty$ and $\gamma \to 0$ is also summarized in \cref{supp-rho-delta-limits-odd-obc}.
    \textbf{(c)} The approximate linear relationship of $\Delta \rho_+^{ss} =\rho_+^{(o)} - \rho_+^{(e)}$ with $m$ under different asymmetry hopping ratios $\gamma$. 
    \tb{(d)} Steady states distinguished by $\Pi$ [\cref{eq:supp-boson-number-parity}] under PBC.
    \tb{(e)} Asymptotic behavior of the steady states under different $\gamma$ for PBC is shown for $\Pi \in \{0,1\}$.
    Parameters used: system size $N=9$, hopping ratio $\gamma=t_+/t_-=10^{-6}$.
   }
    \label{fig:supp-pair-flip-odd-size-ss}
\end{figure}

\paragraph{Symmetry reduction and uniform steady states under PBCs} 
In contrast to the even-$N$ case [\cref{sec:s5}], $m$ is no longer a dynamic invariant under PBCs. It is because the “wrap-around” pair–flip term, $\hat{b}^{\dagger}_{N,\eta}\,\hat{b}_{N,-\eta}\,\hat{b}^{\dagger}_{1,\eta}\,\hat{b}_{1,-\eta}$
which couples the boundary sites $x=N$ and $x=1$, both with odd indices. As a result, this process modifies $m$ by
$
  \Delta m = \Big[(-1)^{N+1} + (-1)^{1+1}\Big]\cdot(\pm1) = \pm2,
$
thus connecting different $m$-sectors and breaking $m$-conservation. Nevertheless, the system preserves a residual $\mathbb{Z}_2$ symmetry associated with the parity of the total number of $+$ bosons, quantified by the \emph{boson-number parity}:
\begin{equation}
\Pi = \bigg(\sum_x n_{x,+}\bigg) \bmod 2 .
\label{eq:supp-boson-number-parity}
\end{equation}
Consequently, the state space fragments into two dynamically disconnected sectors, distinguished by $\Pi \in \{0,1\}$ [\cref{fig:supp-pair-flip-odd-size-ss}(d,e)].
And the steady states corresponding to $\Pi\in\{0,1\}$ are uniform distributions under PBC for odd-sized system, 
\begin{equation}
 \rho_+^{ss}(x)=
    \begin{cases}
      \displaystyle \frac{\sum_{i=0}^l \binom{2l+1}{2i}\gamma^i \frac{2i}{2l+1}}{\sum_{i=0}^l \binom{2l+1}{2i}\gamma^i } & \text{if }\Pi=0, \\
      \displaystyle \frac{\sum_{i=0}^l \binom{2l+1}{2i+1}\gamma^i \frac{2i+1}{2l+1}}{\sum_{i=0}^l \binom{2l+1}{2i+1}\gamma^i} & \text{if }\Pi=1,
    \end{cases}
    \label{eq:supp-pair-flip-ss-odd-pbc}
\end{equation}
where $\gamma:=t_+/t_-$ and $l=\left\lfloor N/2 \right \rfloor$ is half the system size.

\subsection{Derivation of the staggered occupation profile in steady states under OBCs}\label{sec:s6.2}
In this section, we derive the analytic form of the steady state [\cref{eq:supp-pair-flip-ss-odd-obc}] for the correlated spin-flip model under OBCs,  
\[
 L=\sum_{x,\eta=\{\pm\}}t_{\eta}(\hat{n}_{x,-\eta}\hat{n}_{x+1,-\eta}-\hat{b}_{x,\eta}^{\dagger}\hat{b}_{x,-\eta}\hat{b}_{x+1,\eta}^{\dagger}\hat{b}_{x+1,-\eta}),
\]
for odd-sized systems. We define $\gamma:=t_+/t_-$ as the asymmetric hopping ratio and $l= \left \lfloor N/2 \right \rfloor$ as half the system size for convenience.

Under OBCs, the even-odd site imbalance defined in \cref{eq:even-odd-imbalance}, i.e., $m=\sum_{x=1}^{N} (-1)^{x+1} n_{x,+}$
remains a dynamic invariant for odd-sized systems. The analytical derivation of the steady state proceeds analogously to the case with even system size $N$. But the key distinction is that the number of odd-indexed sites is $A = l + 1$, which is \emph{different} from the number of even-indexed sites $B = l$. Similarly, the state space decomposes into $N + 1$ disjoint sectors labeled by $m$, with the dimension of each $m$-sector given by $\mathcal{D}_m = \binom{2l + 1}{l + m}$.

We start with $m>0$. Within a given $m$-sector, basis states can be classified by the number of occupied sites, denoted as $n_{tot} = A + B$, where $A$ and $B$ represent the number of occupied sites at odd and even indices, respectively. 
By analyzing correlated spin-flip    dynamics, $n_{tot}$ starts with $m = A - B$ and varies in steps of 2, such that $n_{tot} = m, m+2, \dots, 2l+1-m$, where $l = \left\lfloor N/2\right\rfloor$. According to the detailed balance condition in the steady state, when a state with $n_{tot}$ occupied sites transitions to one with $n_{tot}+2$ occupied sites (e.g., via a transition $\ket{...,00,...}\to \ket{...,11,...}$), the steady-state wavefuntion satisfies $\frac{\psi_{n_{tot}+2}^{ss}}{\psi_{n_{tot}}^{ss}} = \frac{t_+}{t_-}=\gamma$, where $\psi^{ss}_{n_{tot}}$ is the probability amplitude for basis states with $n_{tot}$. Consider all possible $n_{tot}$ within the $m$ sector, the steady-state wavefunction follows an exponential pattern $[\psi_{n_{tot}=m}, \psi_{n_{tot}=m+2}, ...,\psi_{n_{tot}=m+2i},\dots,\psi_{n_{tot}=2l+1-m}] = \frac{1}{\mathcal{N}_m^+}[1, \gamma, \dots, \gamma^i,\dots, \gamma^{l+1-m}]$. $\mathcal{N}_m^+$ is a normalization constant 
set by the probability conservation condition
\begin{equation}
    \sum_{i=0}^{l+1-m}\binom{l+1}{i+m}\binom{l}{i} \gamma^{i} \frac{1}{\mathcal{N}_m^+} = 1
    \label{eq:prob norm},
\end{equation}
where the occupation numbers at odd and even sites are $m+i$ and $i$ and then $\binom{l+1}{i+m}\binom{l}{i}$ is the degeneracy of states corresponding to specific $m$ and $n_{tot}=m+2i$.

Given $n_{tot}=m+2i$, the occupation numbers at odd and even sites are $m+i$ and $i$ respectively.  
The average occupation at an odd-indexed site is given by $\frac{i+m}{l+1}$, where the number of odd-indexed sites is $l+1$. 
Next, the probability of states with $(n_{tot}=m+2i,m)$ is $\binom{l+1}{i+m}\binom{n}{i}\gamma^i \frac{1}{\mathcal{N}_m^+}$. 
The expected occupation at an odd site is then given by $\binom{l+1}{i+m}\binom{l}{i}\gamma^i (\frac{i+m}{l+1}) \frac{1}{\mathcal{N}_m^+}$. 
Similarly, the expected occupation at even sites is $\binom{l+1}{i+m}\binom{l}{i}\gamma^i (\frac{i}{l}) \frac{1}{\mathcal{N}_m^+}$. 
Substituting \cref{eq:prob norm} into $\frac{1}{\mathcal{N}_m^+}$ and denoting $\gamma=\frac{t_+}{t_-}$, the steady-state occupation profile, $\rho_+^{ss}(x)= \langle n_{x,+}\rangle$, exhibits a staggered pattern under OBC,
\begin{equation}
\rho_+^{ss}(x)=
\begin{cases}
   \rho_+^{(o)} & \text{if }x \text{ is odd,} \\
   \rho_+^{(e)} & \text{if }x \text{ is even,} 
\end{cases}
\end{equation}
with $\rho_+^{(o)}$ and $\rho_+^{(e)}$ determined by $m$ as follows:
\begin{equation}
\begin{cases}
    \rho_+^{(o)} = \sum_{i=0}^{l+1-m}\binom{l+1}{i+m}\binom{l}{i} \gamma^{i} (\frac{i+m}{l+1}) /\mathcal{N}_{m}^+ \\
    \rho_+^{(e)} = \sum_{i=0}^{l+1-m}\binom{l+1}{i+m}\binom{l}{i} \gamma^{i} (\frac{i}{l}) /\mathcal{N}_{m}^+
\end{cases}\text{ if } \ m>0,
\end{equation}
where $\mathcal{N}_{m}^+ = \sum_{i=0}^{l+1-m}\binom{l+1}{i+m}\binom{l}{i} \gamma^{i}$.
A similar analysis also applies to $m\leq0$ case,
\begin{equation}
\begin{cases}
    \rho_+^{(o)} = \sum_{i=0}^{l-|m|}\binom{l+1}{i}\binom{l}{i+|m|} \gamma^{i} (\frac{i}{l+1}) /\mathcal{N}_{m}^- \\
    \rho_+^{(e)} = \sum_{i=0}^{l-|m|}\binom{l+1}{i}\binom{l}{i+|m|} \gamma^{i} (\frac{i+|m|}{l}) /\mathcal{N}_{m}^-
\end{cases}\text{ if } \ m\leq0,
\end{equation}
where $\mathcal{N}_{m}^- = \sum_{i=0}^{l-|m|}\binom{l+1}{i}\binom{l}{i+|m|} \gamma^{i}$.

\subsection{Asymptotic analysis of staggered occupation in the steady state}\label{sec:s6.3}
We now analyze the asymptotic behaviour of the steady-state occupation, i.e., $\rho_+^{(o)}$ and $\rho_+^{(e)}$ [\cref{eq:supp-pair-flip-ss-odd-obc-m<0,eq:supp-pair-flip-ss-odd-obc-m>0} in the following three regimes. 

\tb{1.} $\gamma \to 0$:

For $m>0$, the contributions are dominated by the 0-th order of $\gamma$ in $\rho_+^{(o)}$ and $\rho_+^{(e)}$:
\begin{equation}
    \rho_+^{(o)}\to \frac{m}{l+1} \quad \rho_+^{(e)} \to 0,
\end{equation}
for $m\leq 0$ case, similarly,
\begin{equation}
    \rho_+^{(o)}\to 0 \quad \rho_+^{(e)}\to \frac{|m|}{l},
\end{equation}
And thus, we have
\begin{equation}
\Delta \rho_+^{ss}= \rho_+^{(o)}- \rho_+^{(e)}\xrightarrow[\gamma\to 0]{}
    \begin{cases}
        \frac{m}{l+1}  &  m>0\\
        \frac{-|m|}{l}  & m\leq 0 .
    \end{cases}
\end{equation}

For $m>0$, we have
\begin{equation}
    \mathcal{N}_m^+ = \sum_{i=0}^{l+1-m} \binom{l+1}{i+m}\binom{l}{i} 
    =\sum_{i=0}^{l+1-m} \binom{l+1}{i+m}\binom{l}{l-i}
    = \binom{2l+1}{l+m}, 
\end{equation}
where we use the Vandermonde (Chu–Vandermonde) convolution $\sum_{i=0}^n \binom{A}{i} \binom{B}{k-i} =\binom{A+B}{k}$, which easily follows from tracking the powers of $x$ in $(1+x)^{A+B} = (1+x)^A (1+x)^B$.
Thus \begin{equation}
    \begin{aligned}
        \rho_+^{(o)} &= \frac{\sum_{i=0}^{l+1-m} \binom{l+1}{i+m}\binom{l}{i} \frac{i+m}{l+1}}{\mathcal{N}_m^+} = \frac{\sum_{i=0}^{l+1-m}\binom{l}{i+m-1}\binom{l}{i}}{\binom{2l+1}{n+m}} \\
        &=\frac{\binom{2l}{l+m-1}}{\binom{2l+1}{n+m}} = \frac{l+m}{2l+1}.     
    \end{aligned}
\end{equation}
Similarly, we have 
\begin{equation}
    \begin{aligned}
        \rho_+^{(e)} &= \frac{\sum_{i=0}^{l+1-m}\binom{l+1}{i+m}\binom{l}{i}\frac{i}{l}}{\mathcal{N}_m^+} = \frac{\sum_{i=0}^{l+1-m}\binom{l+1}{i+m}\binom{l-1}{i-1}}{\mathcal{N}_m^+}\\
        &= \frac{\binom{2l}{l+m}}{\binom{2l+1}{l+m}} = \frac{l-m+1}{2l+1}.
    \end{aligned}
\end{equation}

For $m\leq 0$ case, we have
\begin{equation}
    \mathcal{N}_m^- = \sum_{i=0}^{l-|m|} \binom{l+1}{i}\binom{l}{i+|m|} = \binom{2l+1}{l+1+|m|},
\end{equation}
then \begin{equation}
    \begin{aligned}
        \rho_+^{(o)} &= \frac{\sum_{i=0}^{l-|m|} \binom{l+1}{i}\binom{l}{i+|m|} \frac{i}{l+1}}{\mathcal{N}_m^-} = \frac{\sum_{i=0}^{l-|m|} \binom{l}{i-1}\binom{l}{i+|m|}}{\mathcal{N}_m^-}\\
        &= \frac{\binom{2l}{l+1+|m|}}{\binom{2l+1}{l+1+|m|}} = \frac{l-|m|}{2l+1}.
    \end{aligned}
\end{equation}
Similarly, we have \begin{equation}
    \begin{aligned}
        \rho_+^{(e)} &= \frac{\sum_{i=0}^{l-|m|} \binom{l+1}{i}\binom{l}{i+|m|} \frac{i+|m|}{l}}{\mathcal{N}_m^-} = \frac{\sum_{i=0}^{l-|m|} \binom{l+1}{i}\binom{l-1}{i+|m|-1}}{\mathcal{N}_m^-} \\
        &= \frac{\binom{2l}{l+|m|}}{\binom{2l+1}{l+1+|m|}} = \frac{l+|m|+1}{2l+1}.
    \end{aligned}
\end{equation}
Noted that the expression for $m>0$ and $m\leq0$ can be unified as
\begin{align}
    \rho_+^{(o)} &\xrightarrow[\gamma\to 0]{} \frac{l+m}{2l+1}; \\
    \rho_+^{(e)} &\xrightarrow[\gamma\to 0]{} \frac{l-m+1}{2l+1};\\
    \Delta \rho_+^{ss} &= \rho_+^{(o)}-\rho_+^{(e)} \xrightarrow[\gamma\to 0]{} \frac{2m-1}{2l+1}.
\end{align}

\tb{3.} $\gamma \to \infty:$

Under the limit $\gamma\to\infty$, the largest power 
$\gamma^i=\gamma^{l+1-m}$ dominates, and we have
\begin{equation}
\begin{cases}
    \rho^{(o)}_+\to 1 \quad \rho^{(e)}_{+} \to \frac{l+1-m}{l} & \text{if }\  m>0 \\
    \rho^{(o)}_+\to \frac{l-|m|}{l+1} \quad \rho^{(e)}_{+} \to 1 & \text{if }\ m\leq 0 .
\end{cases}
\end{equation}
And thus, 
\begin{equation}
\Delta \rho_+^{ss} \xrightarrow[\gamma\to \infty]{}
    \begin{cases}
        \frac{m-1}{l}  &  m>0\\
        \frac{-(|m|+1)}{l+1}  & m\leq 0 .
    \end{cases}
\end{equation}
It is also instructive to examine the scaling of the even-odd density contrast, $\Delta \rho_+^{ss} = \rho_+^{(o)} - \rho_+^{(e)}$, with system size. We show that $\Delta \rho_+^{ss} \propto \frac{1}{l}$, where $l = \lfloor N/2 \rfloor$ and $N$ is the system size. \\
For $m>0$ case, 
\begin{equation}
\begin{aligned}
    \Delta \rho_+^{ss} &= \frac{1}{\mathcal{N}_m^+} \sum_{i=0}^{l+1-m} \binom{l+1}{i+m}\binom{l}{i} \gamma^i\left( \frac{i+m}{l+1} -\frac{i}{l}\right) \\
    &=\frac{m}{l+1} -\frac{1}{l(l+1)\mathcal{N}_m^+} \sum_{i=0}^{l+1-m} i\cdot \binom{l+1}{i+m}\binom{l}{i} \gamma^i \\
    &=\frac{m}{l+1}-\frac{\gamma \mathcal{N}'^+_m(\gamma)}{l(l+1)\mathcal{N}_m^+(\gamma)} \\
    & \propto \frac{1}{l},
\end{aligned}
\end{equation}
where $\mathcal{N}_{m}^+(\gamma) = \sum_{i=0}^{l+1-m}\binom{l+1}{i+m}\binom{l}{i} \gamma^{i}$. For $m\leq0$ case,
\begin{align}
    \Delta \rho_+^{ss} &= \frac{1}{\mathcal{N}_m^-} \sum_{i=0}^{l-|m|} \binom{l+1}{i}\binom{l}{i+|m|} \gamma^i\left( \frac{i}{l+1} -\frac{i+|m|}{l}\right) \\
    &=-\frac{|m|}{l+1} -\frac{1}{l(l+1)\mathcal{N}_m^-} \sum_{i=0}^{l-|m|} (i+|m|)\cdot \binom{l+1}{i}\binom{l}{i+|m|} \gamma^i \\
    & \propto \frac{1}{l},
\end{align}
where $\mathcal{N}_{m}^-(\gamma) = \sum_{i=0}^{l-|m|}\binom{l+1}{i}\binom{l}{i+|m|} \gamma^{i}$.

In summary, we give the asymptotic steady-state occupation behavior of the correlated spin-flip model for odd-sized systems under OBCs.
\renewcommand{\arraystretch}{2}
\setlength{\tabcolsep}{10pt}
\begin{table}[H]
\centering
\begin{tabular}{c c c c c}
\toprule
Limit & Case & $\rho_+^{(o)}$ & $\rho_+^{(e)}$ & $\Delta \rho_+^{ss}=\rho_+^{(o)}-\rho_+^{(e)}$ \\
\midrule
$\gamma \to 1$      &    & $\displaystyle \frac{l+m}{2l+1}$            & $\displaystyle\frac{l-m+1}{2l+1}$                               & $\displaystyle \frac{2m-1}{2l+1}$          \\
$\gamma \to 0$      & $m>0$   & $\displaystyle \frac{m}{l+1}$            & $0$                               & $\displaystyle \frac{m}{l+1}$          \\
$\gamma \to 0$      & $m\le0$ & $0$                                     & $\displaystyle \frac{-m}{l}$      & $\displaystyle \frac{m}{l}$         \\
$\gamma \to \infty$ & $m>0$   & $1$                                     & $\displaystyle \frac{l+1-m}{l}$    & $\displaystyle \frac{m-1}{l}$           \\
$\gamma \to \infty$ & $m\le0$ & $\displaystyle \frac{l+m}{l+1}$      & $1$                               & $\displaystyle \frac{m-1}{l+1}$     \\
\bottomrule
\end{tabular}
\caption{Asymptotic steady-state occupation behavior for odd $N$-sized systems under OBCs, where $l=\left \lfloor N/2 \right \rfloor$ and $\gamma:=t_+/t_-$. }
\label{supp-rho-delta-limits-odd-obc}
\end{table}

\subsection{Derivation of the uniform steady-state occupation profiles under PBCs}\label{sec:s6.4}
In this section, we derive the analytic form of the steady state [\cref{eq:supp-pair-flip-ss-odd-pbc}] for the correlated spin-flip model under PBCs:  
\[
 L=\sum_{x,\eta=\{\pm\}}t_{\eta}(\hat{n}_{x,-\eta}\hat{n}_{x+1,-\eta}-\hat{b}_{x,\eta}^{\dagger}\hat{b}_{x,-\eta}\hat{b}_{x+1,\eta}^{\dagger}\hat{b}_{x+1,-\eta}),
\]
for systems with odd system size $N$. 
Under PBCs, $m$ is no longer conserved in this case, because both ends of the chain are connected and its sites cannot be labeled into even/odd sites. Instead, the system retains a global $\mathbb{Z}_2$ symmetry associated with the parity of number of $+$ bosons, $\Pi = n_{tot}\bmod 2 \in \{0,1\}$, where $n_{tot}=\left(\sum_x n_{x,+} \right)$. Thus, the state space fragments into two disconnected sectors associated with $\Pi = 0 \text{ or } 1$. Owing to translational invariance and the absence of sublattice imbalance, the steady state is expected to be spatially uniform.
We define $\gamma:=t_+/t_-$ as the asymmetric hopping ratio and $l= \left \lfloor N/2 \right \rfloor$ as half the system size for convenience.

Considering $\Pi=0$ sector first, basis states can be classified by the number of '+' bosons, $n_{tot}=\left(\sum_x n_{x,+} \right)$. Due to the correlated spin-flip dynamics, $n_{tot}$ only varies in steps of 2, such that $n_{tot} = 0, 2, \dots, 2l$, where $l = \left\lfloor N/2\right\rfloor$. According to the detailed balance condition in the steady state, when a state with $n_{tot}$ '+' bosons occupied transitions to one with $(n_{tot}+2)$ '+' bosons occupied (e.g., via a transition $\ket{...,00,...}\to \ket{...,11,...}$), the steady-state wavefuntion satisfies $\frac{\psi_{n_{tot}+2}^{ss}}{\psi_{n_{tot}}^{ss}} = \frac{t_+}{t_-}=\gamma$, where $\psi^{ss}_{n_{tot}}$ is the probability amplitude for basis states with $n_{tot}$.
Consider all basis states within the $\Pi = 0$ sector, the steady-state wavefunction amplitudes follow an exponential form,  
\begin{equation}
    [\psi^{ss}_{n_{tot}=0}, \psi^{ss}_{n_{tot}=2},\psi^{ss}_{n_{tot}=4},...,\psi^{ss}_{n_{tot}=2l}] = \frac{1}{\mathcal{N}_0}[1, \gamma,\gamma^2,..., \gamma^{l}], \label{suppeq:expform}
\end{equation}
where $l=\left\lfloor N/2 \right \rfloor$ and $\mathcal{N}_0$ is the normalization factor. 
Despite the resemblance to a real-space skin effect, the exponential form in \cref{suppeq:expform} instead captures exponential accumulation of up spins in state space, with the fully spin-up state as the boundary.
The degeneracy of states with $n_{tot}=2i$ is given by $\binom{2l+1}{2i}$. The probability normalization condition requires
\begin{equation}
    \sum_{i=0}^{l} \binom{2l+1}{2i} \gamma^i \frac{1}{\mathcal{N}_0} = 1,
\end{equation}
which gives rise to 
\begin{equation}
    \mathcal{N}_0 = \sum_{i=0}^{l}\binom{2l+1}{2i} \gamma^{i}.
    \label{eq:supp-prob-norm-odd-pbc}
\end{equation}
Each configuration state with $n_{tot} = 2i$ contributes an average occupation of $2i / (2l+1)$ per site, with a statistical weight $\binom{2l+1}{2i} \gamma^i / \mathcal{N}_0$. Summing over all such configuration states, the steady-state real-space occupation profile within the $\Pi = 0$ parity sector is
\begin{equation}
 \rho_{+,\Pi=0}^{ss}(x) = \sum_{i=0}^l \binom{2l+1}{2i}\gamma^i \frac{2i}{2l+1} /\mathcal{N}_{0}, \quad 
 \label{eq:supp-pair-flip-ss-odd-pbc-Pi0}
\end{equation}
where $\mathcal{N}_{0} = \sum_{i=0}^l \binom{2l+1}{2i}\gamma^i $. Similarly, in the $\Pi = 1$ sector,
\begin{equation}
 \rho_{+,\Pi=1}^{ss}(x) = \sum_{i=0}^l \binom{2l+1}{2i+1}\gamma^i \frac{2i+1}{2l+1} /\mathcal{N}_{1},
  \label{eq:supp-pair-flip-ss-odd-pbc-Pi1}
\end{equation}
where $\mathcal{N}_{1} = \sum_{i=0}^l \binom{2l+1}{2i+1}\gamma^i $.
In both parity sectors, the steady-state occupation increases monotonically with $\gamma = t_+/t_-$.

\subsection{Asymptotic analysis of uniform steady-state occupations}\label{sec:s6.5}
Next, we analyze the asymptotic behavior of the steady-state occupation distribution of the correlated spin-flip model [\cref{eq:supp-pair-flip-Lap}] under PBC, as given by $\rho_{+,\Pi={0,1}}^{ss}(x)$ in \cref{eq:supp-pair-flip-ss-odd-pbc-Pi0,eq:supp-pair-flip-ss-odd-pbc-Pi1} in the limits $\gamma\to\infty$ and $\gamma\to 0$.
We first present the derivation of $\rho_{+,\Pi=1}^{ss}(x)$ and the case of $\rho_{+,\Pi=1}^{ss}(x)$ is analogous.

We first rewrite $\mathcal{N}_0$ in \cref{eq:supp-pair-flip-ss-odd-pbc-Pi0} using the binomial theorem, 
\begin{equation}
    \begin{aligned}
        \mathcal{N}_0 &= \sum_{i=0}^l \binom{2l+1}{2i} \gamma^i= \frac{1}{2}\left[\sum_{i=0}^{2l+1} \binom{2l+1}{i} \Big( (\sqrt{\gamma})^i +(-\sqrt\gamma)^i\Big)\right] \\
        &= \frac{1}{2} \Big[(1+\sqrt{\gamma})^{2l+1} +(1-\sqrt{\gamma})^{2l+1} \Big].
    \end{aligned}
\end{equation}
The nominator of \cref{eq:supp-pair-flip-ss-odd-pbc-Pi0} can then be recast as 
\begin{equation}
    \begin{aligned}
        \sum_{i=0}^l \binom{2l+1}{2i} \gamma^i \frac{2i}{2l+1} &= \frac{2\gamma}{2l+1} \frac{d}{d\gamma} \mathcal{N}_0(\gamma) \\
        &= \frac{\sqrt{\gamma}}{2} \Big[(1+\sqrt{\gamma})^{2l} -(1-\sqrt{\gamma})^{2l}\Big].
    \end{aligned}
\end{equation}
Hence, \cref{eq:supp-pair-flip-ss-odd-pbc-Pi0} is expressed as 
\begin{equation}
     \rho_{+,\Pi=0}^{ss}(x) = \sum_{i=0}^l \binom{2l+1}{2i}\gamma^i \frac{2i}{2l+1} /\mathcal{N}_{0} 
      =\frac{\sqrt{\gamma} (\alpha^{2l}-\beta^{2l})}{\alpha^{2l+1}+\beta^{2l+1}},
\end{equation}
where we define $\alpha=1+\sqrt{\gamma}$ and $\beta=1-\sqrt{\gamma}$ for compactness.
Finally, the steady-state occupation distributions read
\begin{align}
    \rho^{ss}_{+,\Pi=0}(x) &= \frac{\sqrt{\gamma}(\alpha^{2l}-\beta^{2l})}{\alpha^{2l+1}+\beta^{2l+1}}, \\
    \rho^{ss}_{+,\Pi=1}(x) &= \frac{\sqrt{\gamma}(\alpha^{2l}+\beta^{2l})}{\alpha^{2l+1}-\beta^{2l+1}}.
    \label{eq:rho-pbc-odd}
\end{align}
Notably, these distributions are independent of $x$ as they correspond to uniform real-space distributions. 
Hence, we will omit $x$ from \cref{eq:rho-pbc-odd} and denote them by $\rho^{ss}_{+,\Pi=0,1}$. 
Here $l=\lfloor N/2 \rfloor$ is fixed for a system of size $N$.

We now examine the asymptotic behavior of $\rho^{ss}_{+,\Pi=0,1}$ in the limits $\gamma \to 1$, $\gamma \to \infty$, and $\gamma \to 0$, as well as in the thermodynamic limit $l \to \infty$ with fixed $\gamma \neq 1$.

\tb{1. }$\gamma \to 1$, 

In this regime, $\alpha = 1 + \sqrt{\gamma} \to 2, \quad \beta = 1 - \sqrt{\gamma} \to 0$, then we have
\[
\alpha^{2l} \sim 2^{2l}, \quad \beta^{2l} \sim 0, \quad \alpha^{2l+1} \sim 2^{2l+1}, \quad \beta^{2l+1} \sim 0.
\]
Hence,
\[
\rho_{+,\Pi=0}^{ss} \sim \frac{\sqrt{\gamma}(2^{2l} - 0)}{2^{2l+1} + 0}
 = \tfrac{1}{2}, \quad
\rho_{+,\Pi=1}^{ss} \sim \frac{\sqrt{\gamma}(2^{2l} + 0)}{2^{2l+1} - 0}
 = \tfrac{1}{2}.
\]

\textbf{2. }$ \gamma \to \infty $

In the $\Pi = 0$ sector, for $\gamma \gg 1$, we have
  \begin{equation}
  \begin{aligned}
       \rho^{ss}_{+,\Pi=0} &= \frac{\sqrt{\gamma}(\alpha^{2l}-\beta^{2l})}{\alpha^{2l+1}+\beta^{2l+1}}
       =\frac{\sqrt{\gamma}\left[ \binom{2l}{k}(\sqrt{\gamma})^k-\binom{2l}{k}(-\sqrt{\gamma})^k\right]}{\binom{2l+1}{k}(\sqrt{\gamma})^k+\binom{2l+1}{k}(-\sqrt{\gamma})^k} \\
       &\sim \frac{2l \gamma^n}{(2l+1)\gamma^n}\\
       &\sim \frac{2l}{2l+1}.
       \end{aligned}
  \end{equation}
In the $\Pi = 1$ sector, taking the limit $\gamma \to \infty$, we obtain
\[
 \sqrt{\gamma}(\alpha^{2l} + \beta^{2l}) \sim 2\sqrt{\gamma}\,\gamma^l, \quad
\alpha^{2l+1} - \beta^{2l+1} \sim 2\,\gamma^{l+\tfrac{1}{2}}.
\]
Consequently,
\begin{equation}
    \begin{aligned}
        \rho^{ss}_{+,\Pi=1} &= \frac{\sqrt{\gamma}(\alpha^{2l}+\beta^{2l})}{\alpha^{2l+1}-\beta^{2l+1}}\xrightarrow[\gamma \to \infty]{} \frac{2\sqrt{\gamma}\,\gamma^l}{2\,\gamma^{l+\tfrac{1}{2}}} = 1.
    \end{aligned}
\end{equation}
  
\tb{3.} $ \gamma \to 0 $

Similar as $ \gamma \to \infty $, for $ \gamma \to 0 $, we have
\begin{equation}
   \rho_{+,\Pi=0}^{ss} \to 0, \quad \rho_{+,\Pi=1}^{ss} \to \frac{1}{2l+1}.
\end{equation}

\tb{4. Large system size}, $ l \to \infty $, fixed $ \gamma \neq 1 $

Recall that 
\begin{equation}
\rho_{+,\Pi=0}^{ss} = \frac{\sqrt{\gamma}(\alpha^{2l} - \beta^{2l})}{\alpha^{2l+1} + \beta^{2l+1}}, \quad
\rho_{+,\Pi=1}^{ss} = \frac{\sqrt{\gamma}(\alpha^{2l} + \beta^{2l})}{\alpha^{2l+1} - \beta^{2l+1}}.
\end{equation}

Since $|\beta/\alpha| < 1$, the terms involving $\beta^{2l}$ and $\beta^{2l+1}$ decay exponentially in the large-$l$ limit (with $l$ denoting half the system size):
\begin{equation}
\frac{\beta^{2l}}{\alpha^{2l}} \sim \left|\frac{1 - \sqrt{\gamma}}{1 + \sqrt{\gamma}}\right|^{2l} \to 0.
\end{equation}
Therefore, retaining only the leading contribution, the asymptotic behavior is given by
\begin{equation}
\rho_{+,\Pi=0}^{ss} \to \frac{\sqrt{\gamma} \alpha^{2l}}{\alpha^{2l+1}} = \frac{\sqrt{\gamma}}{\alpha} = \frac{\sqrt{\gamma}}{1 + \sqrt{\gamma}}, \quad
\rho_{+,\Pi=1}^{ss} \to \frac{\sqrt{\gamma}}{1 + \sqrt{\gamma}}.
\end{equation}

The asymptotic form of the uniform steady-state distribution for the correlated spin-flip model with an odd number of sites under PBCs is summarized below:

\newcolumntype{P}[1]{>{\centering\arraybackslash}p{#1}}
\renewcommand{\arraystretch}{2}
\begin{table}[H]
\centering
\begin{tabular}{c c c c}
\toprule
Limit & $\rho_{+,\Pi=0}^{ss}$ & $\rho_{+,\Pi=1}^{ss}$ & $\rho_{+,\Pi=1}^{ss} - \rho_{+,\Pi=0}^{ss}$ \\
\midrule
$\gamma \to 0$      & $0$                                   & $\displaystyle \frac{1}{2l+1}$            & $\displaystyle \frac{1}{2l+1}$ \\
$\gamma \to 1$      & $\tfrac{1}{2}$                        & $\tfrac{1}{2}$                           & $0$ \\
$\gamma \to \infty$ & $\displaystyle \frac{2l}{2l+1}$       & $1$                                      & $\displaystyle \frac{1}{2l+1}$ \\
$l \to \infty$      & $\displaystyle \frac{\sqrt{\gamma}}{1+\sqrt{\gamma}}$ & $\displaystyle \frac{\sqrt{\gamma}}{1+\sqrt{\gamma}}$ & $0$ \\
\bottomrule
\end{tabular}
\caption{Asymptotic steady-state occupations behavior for odd $N$-sized systems under PBCs, where $l=\left \lfloor N/2 \right \rfloor$ and $\gamma:=t_+/t_-$.}
\label{supp-rho-delta-limits-odd-pbc}
\end{table}

\section{Physical Interpretation of the Markov chain Laplacians in this work}\label{sec:s7}
In this section, we elaborate on the physical meaning of the stochastic lattice models in this work, and what they can approximately correspond to in real life.

\subsection{Physical interpretation of interacting HN model (anti-correlated spin-flip   model)}
The Laplacian for the interacting Hatano-Nelson (HN) model [Eq.~(7) in the main text; see \cref{sec:s1} for details], which can also be mapped to our anti-correlated spin-flip   model, is given by
\begin{equation}
  L = \sum_{x} \sum_{\pm} \lambda_{\pm}\, \Bigl( \hat{\rho}_x - \hat{b}_{x\pm1}^\dagger \hat{b}_{x} \Bigr) \left( n_{max}    - \hat{\rho}_{x\pm1} \right).
\end{equation}
The term $ \hat{b}_{x\pm1}^\dagger \hat{b}_{x} $ corresponds to a hopping from site $x$ to a neighboring site $x\pm1$. The transition rate is modulated by the factor $ \lambda_{\pm} \left( n_{max} - \hat{\rho}_{x\pm1} \right) $, which encodes three key features:
\begin{enumerate}
    \item \emph{Biased hopping} — the asymmetry between $ \lambda_+ $ and $ \lambda_- $ drives non-reciprocal transitions in a preferred direction.
    \item \emph{Finite capacity} — each site can only accommodate a finite number of particles $n_{max}$, because any further transition into it is vanishes once $n_{max}$ is reached. 
    \item \emph{Repulsive interaction} — Related to the above, transitions into the target site are suppressed when the latter is near its occupation limit, mimicking effective repulsion.
\end{enumerate}
The term $ \hat{\rho}_x \left( n_{max} - \hat{\rho}_{x\pm1} \right) $ ensures that the total probability is conserved, effectively counteracting any local gain or loss from the the off-diagonal hoppings $\hat{b}_{x\pm1}^\dagger \hat{b}_{x}$.

This model is sufficient for capturing several aspects of the variety of real-life systems. 
Several scenarios are given, for example:
\begin{itemize}
  \item \textbf{Multi-Level Shopping Mall:} 
  Each level of the mall can be identified with a discrete site that has a maximum allowable occupancy, $n_{max}$ due to space and safety regulations. 
  The maximum capacity in this building is then restricted by the total number of allowable occupancy in each level with $N = \sum n_{max}$.
  The biased transition rates ($\lambda_+ \neq \lambda_-$) towards popular levels reflect the preferential flow of people. 
  The exclusion term ensures that as a floor nears its capacity, the effective rate at which additional passengers can arrive is reduced, mimicking the physical limitations of the system. 
  
  Quite often, $\Lambda_\pm$ are dictated by natural footfall dynamics: all other factors being uniform, a mall that has an entrance only at the ground floor often finds an exponentially decreasing number of shoppers the higher one goes (reminiscent of an exponential NHSE profile), unless an efficient elevator system exists. On the other hand, if a specific floor boasts of a crowd-drawing attraction, the effective skin states do not only reside in the bottom floors; but rather the floor with the most popular stores.
  
  \item \textbf{Transport in Constrained Environments:} 
  With some extension to admit branches in the HN chain, constrained environments such as traffic bottlenecks or ions moving through narrow channels in biological membranes can also be captured by the interacting HN model. 
  Given that the maximum total occupancy number in the constrained environment is $N$, the finite capacity, $n_{max}$ can represent the maximum number of agents i.e. vehicles or ions, that can occupy a lane segment or the limited binding sites available in a channel, respectively. 
  This finite capacity also doubles up as the repulsion effect per lane or channel.
  The asymmetry in the hopping rates, $\lambda_+ \neq \lambda_-$ describes external driving forces or gradients (e.g., pressure or voltage differences).

  Indeed, the interacting HN model exhibits key features observed in real-world systems where transport is constrained by local occupancy limits and directional biases significantly influence the steady-state distribution of agents. 
As such, it may serve as a simplified framework for capturing aspects of these complex dynamics.
  
\end{itemize}

\subsection{Physical Interpretation of the correlated spin-flip    model}

The Laplacian for the correlated spin-flip Laplacian [Eq.~(10) in the main text; see \cref{sec:s5,sec:s6} for details] is 
\begin{align}
    L_{t} = \sum_{x} \sum_{\eta=\pm} t_{\eta}\,\Bigl( \hat{n}_{x,-\eta}\, \hat{n}_{x+1,-\eta} - \hat{b}_{x,\eta}^\dagger \hat{b}_{x,-\eta}\, \hat{b}_{x+1,\eta}^\dagger \hat{b}_{x+1,-\eta} \Bigr),
    \label{eq:supp-inter-pair-flip-lap}
\end{align}
where $\hat{b}_{x,\eta}^\dagger$ creates a particle of species $\eta$ at site $x$, and $\hat{n}_{x,\eta}$ is the corresponding number operator. 
There are three essential spin-exchange mechanisms for this Laplacian:
\begin{enumerate}
    \item \emph{Pair-flipping of locally aligned spins:} The pair-flipping term, $\hat{b}_{x,\eta}^\dagger \hat{b}_{x,-\eta}\, \hat{b}_{x+1,\eta}^\dagger \hat{b}_{x+1,-\eta}$ ensures that only locally aligned nearest neighbor spins are flipped. 
    \item \emph{Binary degree of freedom per site:} This binary characteristic is encoded in $\eta =\pm$, where either only spin-up or spin-down states are physically allowed to occupy any site. This implicitly enforce the single capacity constraint in $L_t$ with $n_{i,+} + n_{i,-} = 1$.
    \item \emph{Asymmetrical spin exchange:} The asymmetry between $t_+ \neq t_-$ implies that there is bias in the exchange from up-up to down-down, compared to vice-versa. 
\end{enumerate}
The single capacity constraint of \cref{eq:supp-inter-pair-flip-lap} is sufficient for capturing several aspects of real-life systems with Boolean degrees of freedom. 

A good example is social opinion dynamics, particularly in polarized communities where individuals adopt one of two mutually exclusive stances—such as agreement ($\eta = +$) or disagreement ($\eta = -$)—on a given issue. We consider a one-dimensional system of $N$ agents, indexed by $x \in \{1, \dots, N\}$. The state of an agent is described by the occupation number $\hat{n}_{x,\eta} \in \{0,1\}$, subject to a hard binary constraint $\hat{n}_{x,+} + \hat{n}_{x,-} = 1$, which ensures that each agent holds only one opinion at a time. The term $\hat{b}_{x,\eta}^\dagger \hat{b}_{x,-\eta}\, \hat{b}_{x+1,\eta}^\dagger \hat{b}_{x+1,-\eta}$ represents scenarios where mutual reinforcement or social influence between neighbors lead to an opinion switch. The coefficients $t_\eta$ encode the directional bias in opinion conversion. When these microscopic processes eventually lead to an extensive (macroscopic) state change within the whole system, a collective opinion conversion takes place. 

\end{document}